\documentclass[useAMS,usenatbib]{mn2e}

\usepackage[psamsfonts]{amssymb}
\usepackage[dvips]{graphicx}
\usepackage{amsmath,alltt}                                                                                                                          
\usepackage{multirow}
\usepackage{rotating}
\usepackage{lscape}

\title[Sizes of Globular Clusters in Giant Galaxies]{Globular Cluster Scale Sizes in Giant Galaxies: Orbital Anisotropy and Tidally Under-filling Clusters in M87, NGC 1399, and NGC 5128}
\author[Webb, J. J. et al.]{Jeremy J. Webb$^{1,2}$ \thanks{E-mail: webbjj@mcmaster.ca (JW)}, Alison Sills$^1$, William E. Harris$^1$, Mat\'\i as G\'{o}mez$^3$, \and Maurizio Paolillo$^{4,5,6}$, Kristin A. Woodley$^7$, Thomas H. Puzia$^8$ \\
$^1$ Department of Physics and Astronomy, McMaster University, Hamilton ON L8S 4M1, Canada \\
$^2$ Department of Astronomy, Indiana University, Bloomington IN 47405, USA \\
$^3$ Departamento de Ciencias Fisicas, Facultad de Ciencias Exactas, Universidad Andres Bello, Chile \\
$^4$ Department of Physical Sciences, University of Napoli Federico II, via Cinthia 9, 80126 Napoli, Italy \\
$^5$ INFN - Napoli Unit, Dept. of Physical Sciences, via Cinthia 9, 80126, Napoli, Italy \\
$^6$ Agenzia Spaziale Italiana Science Data Center, Via del Politecnico snc, 00133, Roma, Italy \\
$^7$ University of California, Santa Cruz, University of California Observatories, 1156 High Street, Santa Cruz, CA 95064, USA \\
$^8$ Institute of Astrophysics, Pontificia Universidad Cat\'{o}lica de Chile, Avenida Vicu\~{n}a Mackenna 4860, Macul, 7820436, Santiago, Chile}

\begin{document}

\pagerange{\pageref{firstpage}--\pageref{lastpage}} \pubyear{2043}

\maketitle

\label{firstpage}

\begin{abstract}

We investigate the shallow increase in globular cluster half-light radii with projected galactocentric distance $R_{gc}$ observed in the giant galaxies M87, NGC 1399, and NGC 5128. To model the trend in each galaxy, we explore the effects of orbital anisotropy and tidally under-filling clusters. While a strong degeneracy exists between the two parameters, we use kinematic studies to help constrain the distance $R_\beta$ beyond which cluster orbits become anisotropic, as well as the distance $R_{f\alpha}$ beyond which clusters are tidally under-filling. For M87 we find $R_\beta > 27$ kpc and $20 < R_{f\alpha} < 40$ kpc and for NGC 1399 $R_\beta > 13$ kpc and $10 < R_{f\alpha} < 30$ kpc. The connection of $R_{f\alpha}$ with each galaxy's mass profile indicates the relationship between size and $R_{gc}$ may be imposed at formation, with only inner clusters being tidally affected. The best fitted models suggest the dynamical histories of brightest cluster galaxies yield similar present-day distributions of cluster properties. For NGC 5128, the central giant in a small galaxy group, we find $R_\beta > 5$ kpc and $R_{f\alpha} > 30$ kpc. While we cannot rule out a dependence on $R_{gc}$, NGC 5128 is well fitted by a tidally filling cluster population with an isotropic distribution of orbits, suggesting it may have formed via an initial fast accretion phase. Perturbations from the surrounding environment may also affect a galaxy's orbital anisotropy profile, as outer clusters in M87 and NGC 1399 have primarily radial orbits while outer NGC 5128 clusters remain isotropic.



\end{abstract}



\section{Introduction \label{Introduction}}

The tidal field of a galaxy influences its globular cluster (GC) population by imposing a maximum size that each cluster can reach \citep[e.g.][]{vonhoerner57, king62, innanen83, jordan05, binney08, bertin08, renaud11}. This maximum size is often referred to as the tidal radius $r_t$, the Jacobi radius, or the Roche lobe of the cluster. In all cases, it marks the distance from the cluster at which a star will become unbound as it feels a stronger acceleration towards the host galaxy than it will towards the GC. \citet{vonhoerner57} predicted that:

\begin{equation}\label{rtHoerner}
r_t=r_{gc} \ (\frac{M}{2M_g})^{1/3}
\end{equation}

\noindent for a cluster of mass $M$ on a circular orbit of radius $r_{gc}$, where $M_g$ is the enclosed galactic mass.

Under the assumption that a galaxy can be approximated by an isothermal sphere ($M_g(r_{gc}) \propto r_{gc}$), we expect $r_t \propto r_{gc}^{\frac{2}{3}}$. Since there is no observational evidence that cluster central concentration \textit{c} changes strongly with $r_{gc}$, the mean half-light radius $r_h$ will follow the same scaling relation as $r_t$ \citep{harris96, vdbergh03}. For the Milky Way, which gives us the only cluster population for which we have three dimensional positions and proper motions, we find $r_h \propto r_{gc}^{0.58 \pm 0.06}$ using positions and half-light radii from Harris 1996 (2010 Edition) and proper motions from \citet{dinescu99, dinescu07, dinescu13}. This is a mild but notable discrepancy from the nominal value of $\frac{2}{3}$.

Taking into consideration that only the projected galactocentric distance $R_{gc}$ can be determined for GCs in other galaxies, the relationship between size and distance takes the form $r_t \propto R_{gc}^\alpha$, where $\alpha \sim 0.4 - 0.5$ for typical radial distributions (cluster density $\propto r_{gc}^{-2}$). However, observations in other galaxies appear to disagree with theoretical predictions. From a study of six giant elliptical galaxies, \citet{harris09a} found the combined dataset was best fitted by an $\alpha$ of 0.11. This value is in agreement with observational studies of NGC 4594 ($\alpha$ = $0.19 \pm 0.03$ \citep{spitler06, harris10a}), NGC 4649 ($\alpha$ = $0.14 \pm 0.06$, \citep{strader12}) M87 ($\alpha$ = $0.14 \pm 0.01$, \citep{webb13b}), NGC 4278 ($\alpha$ = $0.19 \pm 0.02$, \citep{usher13}) and NGC 1399 ($\alpha$ = $0.13 \pm 0.03$ (data from \citet{puzia14})). Looking at the metal poor (blue) and metal rich (red) cluster sub-populations in NGC 5128 separately, \citet{gomez07} found $\alpha = 0.05 \pm 0.05$ for the blue clusters and $\alpha = 0.26 \pm 0.06$ for the red clusters. Only the cluster population of the giant elliptical galaxy NGC 4365 in the Virgo cluster has a measured $\alpha$ of $0.49 \pm 0.04$ that is comparable to the expected range of 0.4-0.5 \citep{blom12}, which may indicate the galaxy has a different dynamical age or has undergone a different formation scenario than the galaxies listed above.

The discrepancy between Equation \ref{rtHoerner} and observed values of $\alpha$ may be attributed to assuming that all GCs have circular orbits in spherically symmetric isothermal tidal fields and that they fill their theoretical $r_t$. The first assumption is required in order for the tidal field experienced by the cluster to be static. However, galaxies will not necessary have isothermal mass profiles or be spherically symmetric. A non-isothermal mass profile will alter the expected value of $\alpha$, with a more strongly increasing cumulative mass with radius ($\frac{d(log(M_g(r_{gc})))}{d(log(r_{gc}))} > 1$) resulting in smaller values of $\alpha$. Furthermore, no known GC has a truly circular orbit \citep{dinescu99,dinescu07,dinescu13}. Eccentric orbits then subject the cluster to tidal heating and tidal shocks which can provide outer stars enough energy to escape the cluster and energize inner stars to larger orbits \citep[e.g][]{kupper10,renaud11, webb13a, kennedy14}. Clusters on eccentric orbits are also able to re-capture temporarily un-bound stars since the cluster's instantaneous $r_t$ is also time-dependent. $N$-body models of GC evolution have shown that despite spending the majority of their lifetimes at apogalacticon, clusters with eccentric orbits lose mass at a faster rate \citep{baumgardt03} and appear smaller \citep{webb13a} than clusters with circular orbits at apogalacticon. Hence incorporating the effects of orbital eccentricity on cluster evolution could reduce the discrepancy between theoretical and observed values of $\alpha$. The situation will be complicated further if the cluster has an inclined orbit in a non-spherically symmetric potential \citep{madrid14, webb14} or if the cluster has been accreted by the host galaxy via a satellite merger such that its current orbit does not reflect the tidal field in which it formed and evolved \citep{miholics14, bianchini15, renaud15, miholics16}.

The second assumption, that all clusters fill their theoretical $r_t$, we now understand is also unrealistic. While a GC will naturally expand due to two-body interactions \citep{henon61}, it is possible that certain clusters formed compact enough or expand slowly enough such that they have yet to \text{reach} the point of filling their $r_t$ and effectively evolve in isolation. Observationally for such clusters, their limiting radius $r_L$ (the radius at which the cluster's density falls to zero) is less than $r_t$. Observations of Galactic GCs have shown that only approximately $\frac{1}{3}$ of the population are tidally filling, in the sense that $r_L \sim r_t$ \citep{gieles11}. The remaining clusters in the Milky Way are still in the expansion phase and are considered to be tidally under-filling. Under-filling clusters have also been found in NGC 4649, where \citet{strader12} found no evidence for tidal truncation for clusters beyond 15 kpc and in NGC 1399, where \citet{puzia14} found no evidence for truncation beyond 10 kpc. \citet{alexander13} were able to reproduce the observed size distribution of Galactic GCs by assuming that all clusters form initially compact and then expand naturally via two-body interactions until they become tidally filling. After 12 Gyr of evolution, inner clusters which experience a strong tidal field and have small tidal radii have expanded to the point of being tidally filling. Outer clusters, with large tidal radii, still remain tidally under-filling after 12 Gyr since the outer tidal field is weak. Allowing clusters to become more under-filling with increasing $r_{gc}$ offers a second explanation as to why observed values of $\alpha$ are noticeably less than theoretical predictions.

Understanding how the factors discussed above can influence $\alpha$ allows us to use the size distribution of GC populations to constrain many properties of their host galaxy, including its mass and orbital anisotropy profiles. In two previous studies of the giant elliptical galaxy M87 \citep{webb12, webb13b}, we explored the effects of orbital anisotropy and tidal filling on its GC population out to 110 kpc. We found that it was possible to reproduce the observed relationship between $r_h$ and $R_{gc}$ in M87 by allowing cluster orbits to be preferentially radial. However, the degree of radial anisotropy required to reproduce the size distribution of inner and outer region clusters was quite different. This discrepancy was partially minimized by allowing orbital anisotropy to change with $r_{gc}$, but the degree of radial anisotropy in the outer regions of M87 was still much higher than kinematic studies suggested \citep{cote01, strader11}. We were also able to match theory and observations by allowing all clusters to be under-filling, but we only explored the effects of clusters being under-filling by the same amount at all $r_{gc}$.

Given that clusters form with some initial size distribution and can be found over large ranges in $R_{gc}$, clusters can under-fill their $r_t$ either because the local tidal field is weak or they formed extremely compact. The situation is complicated even further by the possibility that some clusters in a galaxy may be accreted. If a cluster that was tidally filling in its original host galaxy is accreted and ends up with an orbit at a large $R_{gc}$, it can appear to be extremely under-filling. So instead orbital anisotropy and tidal filling are likely to be functions of $r_{gc}$ \citep[e.g.][]{cote01,prieto08, zait08, weijmans09, gnedin09, ludlow10, kruijssen12, alexander13}. The next step is to then incorporate these two parameters into our model as functions of $r_{gc}$. 

In this study, we consider the combined effects of orbital anisotropy and tidal filling on GC populations in the giant galaxies M87, NGC 5128, and NGC 1399. Since we are focused on giant elliptical galaxies which are spherically symmetric over the range of $R_{gc}$ that our observational datasets cover, orbital inclination is not a contributing factor. It should be noted that some studies have found that M87 is not spherically symmetric at larger $R_{gc}$ and that its surface brightness profile eventually reaches an ellipticity of 0.4 \citep[e.g][]{kormendy09}. Since this ellipticity is not high enough to significantly alter the behaviour of GCs in M87, we can still assume M87 is spherically symmetric in agreement with the many studies which have measured its mass profile \citep{mclaughlin99, strader11, agnello14, zhu14}. However, inclination will have to be considered in future studies if our approach is to be applied to non-spherically symmetric elliptical galaxies and disk galaxies. We also assume that all clusters in a given population have spent their entire lifetimes in the host galaxy. \citet{miholics14} showed that after a cluster is accreted by a host galaxy its size responds to its new potential within 1-2 GC relaxation times and evolves as if it has always orbited in the host galaxy. So while accreted clusters may maintain a kinematic signature of the accretion process, their \textit{structural} parameters (which is the focus of this study) will reflect their \textit{current} orbit in the host galaxy. 

In Section \ref{stwo} we introduce the three observational datasets used in our study and in Section \ref{sthree} we re-introduce the theoretical model used to reproduce the observations. In Section \ref{sfour} we first study how allowing the orbital anisotropy and tidal filling properties of clusters to change with $r_{gc}$ affects the distribution of cluster sizes in each galaxy. We then explore the degeneracy between these two factors and make use of previous kinematic studies of each population to constrain our models even further. The best fit theoretical model for each galaxy is then discussed and the results of all three galaxies are compared in Section \ref{sfive}. We summarize our findings in Section \ref{ssix}.

\section{Observations} \label{stwo}

In the following sections we summarize the datasets for M87, NGC 1399, and NGC 5128 used in this study. Table \ref{table:obs} lists the total number of GCs, radial range, and the measured value of $\alpha$ for each dataset. $\alpha$ is the slope of a log-log plot of median $r_h$ versus $R_{gc}$, where the median $r_h$ is calculated within radial bins containing $5\%$ of the total cluster population. We also list the same properties when clusters are split into red and blue sub-populations. Red and blue clusters in M87 and NGC 1399 have very similar values of $\alpha$, with the global M87 population having a slightly higher value of $\alpha$ overall. Red and blue clusters in NGC 5128 on the other hand have very different vales of $\alpha$, with the $r_h$ of red clusters increasing steeply with $R_{gc}$ compared to blue clusters. Overall, the increase in $r_h$ with $R_{gc}$ is much steeper in NGC 5128 than in M87 or NGC 1399.

\begin{table}
  \caption{Properties of Observed Globular Cluster Populations}
  \label{table:obs}
  \begin{center}
    \begin{tabular}{lcccc}
      \hline\hline
      {Galaxy} & {Population} & {N} & {Radial Range} & {$\alpha$} \\
      \hline
M87 & All & 2335 & 0.1-108 kpc & $0.13 \pm 0.01$ \\
{} & Red & 1211 & 0.1-106 kpc & $0.10 \pm 0.01$ \\
{} & Blue & 1124 & 0.3-108 kpc & $0.10 \pm 0.01$ \\

NGC 1399 & All & 1266 & 2-49 kpc & $0.09 \pm 0.02$ \\
{} & Red & 681 & 2-49 kpc & $0.09 \pm 0.02$ \\
{} & Blue & 584 & 2-48 kpc & $0.08 \pm 0.02$ \\

NGC 5128 & All & 588 & 1.2-47 kpc & $0.19 \pm 0.03$ \\
{} & Red & 310 & 1.2-43 kpc & $0.29 \pm 0.03$ \\
{} & Blue & 278 & 1.4-47 kpc & $0.03 \pm 0.04$ \\
      \hline\hline
    \end{tabular}
  \end{center}
\end{table}

While the colour and luminosity ranges of the M87 dataset are both thoroughly covered down to very low luminosity, in both NGC 1399 and NGC 5128 the data are incomplete for GCs much fainter than the luminosity function turnover. However, as we discuss in Sections \ref{sec:n1399} and \ref{sec:n5128}, the incompleteness in mass range is factored into our model. The main parameter extracted from each dataset is the GC half-light radius, which is measured by fitting the surface brightness profile of each cluster with a \citet{king62} profile. For M87 \citep{webb13b} and NGC 5128 \citep{gomez07, woodley10b}, surface brightness profile fits are done using the commonly used tool ISHAPE \citep{larsen99} and allowing the central concentration to be variable. For NGC 1399, GALFIT \citep{peng10} is used to fit GC surface brightness profiles with \citet{king62} models \citep{puzia14}. $r_h$ has been shown to be a robust parameter that can repeatedly be recovered using different surface brightness profile models \citep{webb12} and different fitting routines \citep{webb13a, puzia14} (including ISHAPE and GALFIT) when cluster sizes are comparable to the point spread function. Therefore using different fitting routines to measure cluster sizes in NGC 1399 will have a minimal affect on the results of this study, especially since our results only rely on the relative trends within each galaxy. Additionally, even though cluster sizes are measured in different wave bands, many studies have found that there are minimal differences when comparing sizes measured with different filters \citep[e.g.][]{harris10b, strader12, webb13b}. And, since similar constraints are used to confirm GC candidates and minimize contaminants (magnitude, colour, quality of fit, and size), the observational datasets are as homogeneous as possible.

\subsection{M87}

M87 is a giant elliptical galaxy located at the centre of the Virgo cluster, with a distance modulus of $(m-M)_0 = 30.88$ \citep{pierce94, tonry01}. Hubble Space Telescope (\textit{HST}) Advanced Camera for Surveys (ACS) / Wide Field Camera (WFC) images of the central 12 kpc of M87 in the F814W (\textit{I}) and F606W (\textit{V}) filters are taken from program GO-10543 (PI Baltz). The more recently completed program GO-12532 (PI Harris) provided a combination of 8 ACS and WFC3 fields of view in the F814W and F475W filters of the outer regions of M87 ranging from 10 kpc to 110 kpc. See \citet{webb12} and \citet{webb13b} for a detailed description of how cluster candidates are selected and sizes are measured.

\subsection{NGC 1399}

NGC 1399 is a giant elliptical galaxy located at the centre of the Fornax cluster, with a distance modulus of $(m-M) = 31.52$ \citep{dunn06,blakeslee09}. In this study we utilize archival \textit{HST} images of NGC 1399 from program GO-10129 (PI Puzia). The 3 $\times$ 3 ACS mosaic in the F606W filter covers approximately 10' x 10 ' out to a projected distance of approximately 50 kpc. A description of how cluster candidates are selected and how sizes are measured can be found in \citet{puzia14}. It should be noted that while \citet{puzia14} introduced a cluster size correction function based on artificial cluster experiments, we use the uncorrected cluster sizes to keep the observational datasets in this study as homogeneous as possible. Hence the quoted value of $\alpha$ for NGC 1399 in Table \ref{table:obs} differs slightly from \citet{puzia14}.

\subsection{NGC 5128}

NGC 5128 (Cen A) is a giant galaxy that is found in relative isolation, with a distance modulus of $(m-M) = 27.92$ \citep{harris10b}. We make use of Magellan/IMACS images of NGC 5128 to study its GC population out to a projected distance of approximately 40 kpc \citep{gomez07}.  A description of how cluster candidates are selected and how sizes are measured can be found in both \citet{gomez07} and \citet{woodley10a}.

\section{Model} \label{sthree}

Our model (first introduced in \citet{webb12} and modified in \citet{webb13b}) generates a mock GC population that has the same distributions in projected distance, velocity and mass as the observed dataset. The model also has the capability to model sub-populations with different radial profiles and velocity dispersions separately, which we apply to the red and blue sub-populations in each galaxy. The central concentration distribution and mass to light ratios of model clusters are set equal to the Milky Way cluster population. Since our model has been modified to be applicable to any galaxy, we will re-introduce it here.

The projected radial distribution of clusters in each galaxy is obtained by fitting the observed number density profile ($n(R_{gc})$) with a modified two-dimensional Hubble profile:

\begin{equation}\label{hubble}
n(R_{gc}) = \frac{n_0}{1.0+(\frac{R_{gc}}{R_0})^2}
\end{equation}

The red and blue radial distributions of GCs in M87 have previously been shown to follow a Hubble profile \citep{harris09b}, and fitting Equation \ref{hubble} to all three of our observed datasets yields reduced $\chi^2$ ($\chi^2_{\nu}$) values of order unity (see Sections \ref{sec:m87} - \ref{sec:n5128} for the best fit values of $n_0$ and $R_0$ for red and blue clusters in each galaxy). Assuming the two-dimensional Hubble profile continues beyond the range of our observational datasets, Equation \ref{hubble} is then transformed to obtain the three dimensional radial distribution from which cluster positions are randomly sampled \citep{binney08}. Each model cluster is then assigned a three dimensional velocity based on the observed global line-of-site velocity dispersion. Before assigning velocities, we first consider the anisotropy parameter ($\beta$), which is one of our two free parameters and controls the degree of orbital anisotropy within the GC system. $\beta$ is defined as \citep{binney08}:

\begin{equation}\label{beta}
\beta =1-\frac{\sigma_\theta^2+\sigma_\phi^2}{2 \sigma_r^2}
\end{equation}

\noindent where $\sigma_r$, $\sigma_\theta$, and $\sigma_\phi$ are the velocity dispersions for each spherical coordinate. In all cases, $\sigma_\theta$ and $\sigma_\phi$ are assumed to be equal. The isotropic case ($\beta=0$) means that $\sigma_r = \sigma_\theta = \sigma_\phi$ are all equal to the line-of-site velocity dispersion and velocities are randomly drawn from a Gaussian distribution. If $\beta$ increases from zero then $\sigma_r$ increases while $\sigma_\theta$ and $\sigma_\phi$ decrease such that the average of all three values still matches the observations and orbits become preferentially radial. The opposite occurs if $\beta$ decreases from the isotropic case and orbits become preferentially circular. 

We also allow for $\beta$ to change as a function of $r_{gc}$. We assume the orbital anisotropy profile of a given galaxy is of the form:

\begin{equation} \label{beta_prof}
\beta(r_{gc}) = \frac{1}{1+(\frac{R_\beta}{r_{gc}})^2}
\end{equation}

where the anisotropy radius $R_\beta$ replaces $\beta$ as one of the free parameters in our model. This form of $\beta(r_{gc})$ is in agreement with theoretical and observational studies which find that inner cluster orbits are primarily isotropic while orbits become preferentially radial with $R_{gc}$ \citep[e.g.][]{cote01, fall01, vesperini03, gnedin09, ludlow10, kruijssen12}. We also note that other functional forms of Equation \ref{beta_prof} have been suggested in the literature, but studying the effects of different $\beta(r_{gc})$ profiles is beyond the scope of this study and will be addressed in the future.

Once a value of $\beta$ has been given to each cluster, a mass is assigned based on the observed luminosity function of the dataset. The mass to light ratio of the model clusters is assumed to be equal to the mean value of $(\frac{M}{L})_V = 2$ found by \citet{mclaughlin05} for Milky Way GCs. The central concentration (\textit{c}) of each cluster (log of the ratio between the cluster's core radius $r_c$ to its limiting radius ($log(\frac{r_c}{r_L})$) is assigned based on the distribution of Milky Way GCs \citep{harris96}, which is Gaussian with a mean of \textit{c}$=1.5$ and dispersion of 0.4.

\begin{figure}
\centering
\includegraphics[width=\columnwidth]{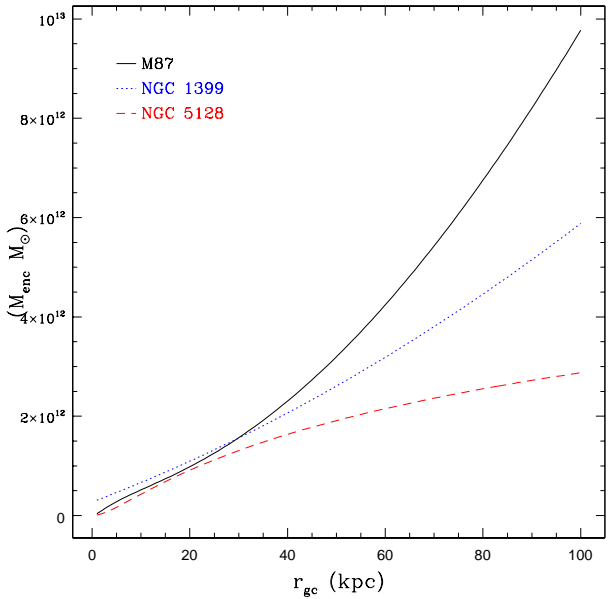}
\caption{ Total enclosed mass as a function of $r_{gc}$ for M87 (black), NGC 1399 (blue), and NGC 5128 (red).}
  \label{fig:mass}
\end{figure}

Once each model cluster has been generated, the mass profile of the selected galaxy (see Figure \ref{fig:mass} and Sections \ref{sec:m87}- \ref{sec:n5128}) is used to calculate the theoretical value of $\alpha$ that is expected assuming all clusters have circular orbits. Knowing the gravitational potential field also allows for the orbit of each individual cluster to be solved \citep{binney08}. The resulting distribution of orbital eccentricities (and its dependence on $r_{gc}$) will then be dependent on $\beta$, the velocity dispersion of the GC population and the mass profile of the host galaxy. Hence two galaxies can have similar values of $\beta$ but different distributions of cluster orbits.

Using the formalism of \citet{bertin08}, we next calculate each cluster's $r_t$ at perigalacticon $r_p$. For clusters with eccentric orbits we use their orbital frequency $\Omega$ at $r_p$ to calculate $r_t$ as opposed to $\Omega=((d\Phi_G(r)/dr)_{r_p}/r_p)^{\frac{1}{2}}$ which assumes the cluster has a circular orbit at $r_p$ \citep{moreno14}. We then determine $r_L$ at $r_p$ based on our second free parameter $R_f=\frac{r_L}{r_t}$, also known as the tidal filling parameter. Hence $R_f$ is a measure of how filling a cluster is at $r_p$, with clusters that fill only a fraction of their permitted $r_t$ having $R_f < 1 $.

We next assume that each model cluster can be represented by a \citet{king62} model, such that the limiting radius at $r_p$ and the previously assigned central concentration set the cluster's surface brightness profile. However, $R_f$, $r_L$ and $r_h$ corresponding to the surface brightness profile are only valid when the cluster is located at $r_p$ and will change as a function of orbital phase. Tidal heating, tidal shocks at $r_p$ and the recapturing of temporarily unbound stars as a cluster moves away from $r_p$ will cause $r_h$ and $r_L$ to increase as a function of orbital phase. We therefore correct the model $r_h$ values for orbital eccentricity following \citet{webb13a} and \citet{webb13b}, with $r_h$ increasing by a maximum of $30\%$ for highly eccentric clusters. No corrections are necessary for $r_L$ and $R_f$ since we do not compare these values to observations.

Similar to $\beta$, $R_f$ can also be a function of a cluster's location in the tidal field. We specifically allow $R_f$ to change as a function of $r_{gc}$ via:

\begin{equation} \label{rf_prof}
R_f(r_{gc}) = 1-\frac{1}{1+(\frac{R_{f\alpha}}{r_{gc}})^2}
\end{equation}

\noindent where the filling radius $R_{f\alpha}$ becomes the free parameter. This form of $R_{f}$ ensures clusters become less tidally filling as the tidal field becomes weaker \citep{alexander13}.

Finally, to best match the observed datasets, we apply magnitude and size cutoffs to the simulated dataset such that the model does not produce GCs that may exist but would not be observed. We also check to make sure the simulation does not produce any clusters with evaporation or infall times due to dynamical friction less than any observed clusters. 

The individual input parameters and mass profiles of each galaxy are discussed in Sections \ref{sec:m87} - \ref{sec:n5128}. Since each galaxy has multiple estimations of various input parameters, we use the best available data that are also in line with the assumptions made by our model. Perhaps the most influential input parameter is our choice of mass profile. While a complete study of the effects that different mass profiles will have on our results is beyond the scope of our study, we note a different rate of increase in mass with $r_{gc}$ will result in clusters having different values of $r_t$. More specifically a steeper increase in enclosed mass with $r_{gc}$ will result in a shallower increase in $r_t$ with $r_{gc}$. If the true values of $r_t$ are smaller than the values calculated using the adopted mass profiles, then GCs must be more tidally filling and/or have a lesser degree of radial anisotropy. Hence our estimation of $R_f$ will be a lower limit and our estimation of $\beta$ will be a upper limit. The opposite will be true if the mass profile is shallower.

\subsection{M87}\label{sec:m87}

The radial profile and luminosity function of our M87 dataset are listed in Table \ref{table:m87sim}, along with the velocity dispersion parameters assigned to our theoretical cluster population. While observations of inner clusters sample the entire luminosity function, we incorporate into our model the fact that the luminosity function of outer clusters is only $\sim 50 \%$ complete beyond the luminosity function turnover. In a kinematic study of the GC population of M87, \citet{cote01} found that blue clusters have a mean velocity (minus the galaxy's systemic velocity) of -36 km/s with a dispersion of 412 km/s while red clusters have a mean velocity of 7 km/s and a dispersion of 385 km/s. They also suggested that the velocity dispersion may increase with $R_{gc}$. However more recent studies by \citet{strader11} and \citet{zhu14} found that the global velocity dispersion stays relatively constant with $R_{gc}$. Due to the larger datasets of \citet{strader11} and \citet{zhu14} and their more rigorous treatment of outliers, we will assume the velocity dispersion of M87 is constant at all $r_{gc}$. There is also no evidence in any of the galaxies presented in this study that either the mean velocity or velocity dispersion is dependent on cluster luminosity. 

The galactic potential of M87 is taken directly from \citet{mclaughlin99} and has the form:

\begin{equation} \label{mass_m87}
M_{total}(r)= M_{stars}(r) + M_{dark}(r)
\end{equation}
\begin{equation}\label{ms}
M_{stars}(r) = 8.10 \times 10^{11} \ M_{\odot} \ [\frac{(r/5.1 \ \mathrm{kpc})}{(1+r/5.1 \ \mathrm{kpc})}]^{1.67}
\end{equation}
\begin{equation} \label{md}
M_{dark}(r)= 7.06 \times 10^{14} \ M_{\odot} \times[\ln(1+r/560 \ \mathrm{kpc})-\frac{(r/560 \ \mathrm{kpc})}{(1+r/560 \ \mathrm{kpc})}]
\end{equation}

The stellar mass component (Equation \ref{ms}) was determined by fitting model mass density profiles for spherical stellar systems \citep{dehnen93, tremaine94} to $B$-band photometry \citep{devaucouleurs78}, assuming the stellar mass-to-light ratio of M87 is independent of radius. The dark matter component of M87 (Equation \ref{md}) was determined by combining X-ray observations of hot gas in the extended M87 halo, dwarf elliptical galaxies, and early-type Virgo galaxies to generate a Navarro-Frenk-White (NFW) dark matter halo \citep{navarro97}. 

The overall mass profile is in general agreement with the more recent kinematic study of M87 performed by \citet{strader11}, though the latter found evidence for a larger dark matter component within 20 kpc. Using the \citet{strader11} dataset, \citet{agnello14} also derived stellar and dark matter mass profiles for M87 by separating its cluster population into three sub-populations and noting their distinct radial distributions and velocity dispersions as a function of $R_{gc}$. The total mass of M87 as determined by \citet{agnello14} is comparable to \citet{mclaughlin99}, although \citet{agnello14} found a more gradual increase in dark matter mass than \citet{mclaughlin99}. Using an even larger GC dataset than \citet{agnello14} over a wider range of $R_{gc}$, \citet{zhu14} found a lower total mass within 100 kpc than \citet{mclaughlin99}. However the gradient in the mass profiles of \citet{mclaughlin99} and \citet{zhu14}, which is the key factor in setting how $r_t$ increases with $r_{gc}$, are very similar out to an $R_{gc}$ of 80 kpc. As noted by \citet{strader11}, more extensive modelling of M87 and Virgo is required in order to better constrain its dark matter halo. Since the differences between the mass profiles discussed above are minimal (within the $R_{gc}$ range of our observed dataset) and the \citet{mclaughlin99} model incorporates X-ray observations, we will utilize the mass profile as determined by \citet{mclaughlin99}.

\begin{table}
  \caption{Simulated M87 Globular Cluster Population Input Parameters}
  \label{table:m87sim}
  \begin{center}
    \begin{tabular}{lc}
      \hline\hline
      {Parameter} & {Value} \\
      \hline
Number of Clusters & \\
Blue & 1124 \\
Red & 1211 \\
Radial Distribution & Modified Hubble Profile \\
Blue Population & \\
$\sigma_0$ & 37.95 arcmin$^{-2}$ \\
$R_0$ & 1.08' \\
Red Population & \\
$\sigma_0$ & 95.7 arcmin$^{-2}$ \\
$R_0$ & 0.83' \\
Angular Distribution & Spherically Symmetric \\
Luminosity Function & Gaussian \\
$\langle M_V \rangle$ & -7.6 \\
$\sigma_{M_V}$ & 1.0 \\
Velocity Dispersion & \citet{cote01} \\
Blue Population & \\
$\langle v \rangle$ & -36 km/s \\
$\sigma_v$ & 412 km/s \\
Red Population & \\
$\langle v \rangle$ & 7 km/s \\
$\sigma_v$ & 385 km/s \\

      \hline\hline
    \end{tabular}
  \end{center}
\end{table}

\subsection{NGC 1399}\label{sec:n1399}

The radial profile and luminosity function of our NGC 1399 dataset are listed in Table \ref{table:n1399sim}. Note that we have incorporated into our model that the luminosity function of our dataset is only complete to an absolute magnitude of -5.7. The velocity dispersion parameters assigned to our theoretical cluster population are taken from the most recent kinematic study of the NGC 1399 GC population, where \citet{schuberth10} found that blue clusters have a mean velocity of 11 km/s with a dispersion of 358 km/s and the red clusters have a mean velocity of 31 km/s with a dispersion of 256 km/s. While \citet{schuberth10} also suggested that the velocity dispersion of red and blue clusters may change with $R_{gc}$, we will assume these values remain constant with $R_{gc}$ to stay consistent with our model for M87.

\citet{schuberth10} also derived a mass profile for NGC 1399 based on GC kinematics, although the authors assumed a value for the anisotropy parameter $\beta$. To stay consistent with the mass profile used for M87, we take the galactic potential of NGC 1399 as derived from \textit{ROSAT} High Resolution Imager data by \citet{paolillo02}. X-ray emission from hot gas in NGC 1399 was used to make enclosed total mass (stars and dark matter) estimates at various distances. Since we require a functional form for the mass profile of each galaxy in order to solve the orbits and calculate the size of each model cluster, and since the data does not reflect a standard NFW profile, we fit the mass profile from \citet{paolillo02} with a quadratic function:

\begin{equation} \label{mass_n1399}
M_{tot}(r)= 2.74 \times 10^{11} \ M_{\odot} + 3.73 \times 10^{10} \ M_{\odot} (\frac{r}{1 kpc})+ 1.87 \times 10^8 M_{\odot} (\frac{r}{1 kpc})^2
\end{equation}
 
\begin{table}
  \caption{Simulated NGC 1399 Globular Cluster Population Input Parameters}
  \label{table:n1399sim}
  \begin{center}
    \begin{tabular}{lc}
      \hline\hline
      {Parameter} & {Value} \\
      \hline
Number of Clusters & \\
Blue & 558 \\
Red & 668 \\
Radial Distribution & Modified Hubble Profile \\
Blue Population & \\
$\sigma_0$ & 17 arcmin$^{-2}$ \\
$R_0$ & 1.1' \\
Red Population & \\
$\sigma_0$ & 50 arcmin$^{-2}$ \\
$R_0$ & 0.8' \\
Angular Distribution & Spherically Symmetric \\
Luminosity Function & Gaussian \\
$\langle M_V \rangle$ & -7.3 \\
$\sigma_{M_V}$ & 1.3 \\
Velocity Dispersion & \citet{schuberth10} \\
Blue Population & \\
$\langle v \rangle$ & 11 km/s \\
$\sigma_v$ & 358 km/s \\
Red Population & \\
$\langle v \rangle$ & 31 km/s \\
$\sigma_v$ & 256 km/s \\

      \hline\hline
    \end{tabular}
  \end{center}
\end{table}

\subsection {NGC 5128}\label{sec:n5128}

The radial profile and luminosity function of our NGC 5128 dataset are listed in Table \ref{table:n5128sim}. We have incorporated into our model the fact that our NGC 5128 cluster dataset is only $60\%$ complete fainter than the luminosity function turnover. The velocity dispersion parameters assigned to our theoretical cluster population (also in Table \ref{table:n5128sim}) are taken from \citet{woodley10b}. In a kinematic study of over 600 GCs, they determined that blue GCs have a mean velocity of 26 km/s with a dispersion of 149 km/s and red clusters have a mean velocity of 43 km/s with a dispersion of 156 km/s. Similar to NGC 1399, \citet{woodley10b} found evidence that the velocity dispersion of red and blue clusters changes with $R_{gc}$. However with no quantitative analysis of this radial variation, we again assume these values remain constant with $R_{gc}$. It is interesting to note that the velocity dispersions of the NGC 5128 cluster populations are approximately a factor of two smaller than in M87 and NGC 1399. This difference must have to do with M87 and NGC 1399 being massive galaxies located at the centres of a rich galaxy cluster while NGC 5128 is more or less in isolation. We will discuss the impact of environment further in Section \ref{sfive}.

A mass profile of NGC 5128 that uses X-ray emission from hot gas currently does not exist. Instead we take the potential of NGC 5128 from enclosed total mass estimates from \citet{woodley10b} based on the kinematics of the NGC 5128 cluster population, which makes assumptions regarding the anisotropy profile of the cluster population. While we note that the mass profile of NGC 5128 has been determined via a different method than M87 and NGC 1399, it is in agreement with previous estimates taken from studies of HI gas shells \citep{schiminovich94}, planetary nebulae \citep{woodley07,peng04a} and other cluster datasets \citep{peng04b} and will still accurately reflect the true $M_{tot}(r)$. Fitting the total mass estimates with a NFW profile \citep{navarro97}, we find:

\begin{equation} \label{mass_n5128}
M_{tot}(r)= 1.74 \times 10^{14} \ M_{\odot} \times[\ln(1+r/8.2 \ \mathrm{kpc})-\frac{(r/8.2 \  \mathrm{kpc})}{(1+r/8.2 \  \mathrm{kpc})}]
\end{equation}

\begin{table}
  \caption{Simulated NGC 5128 Globular Cluster Population Input Parameters}
  \label{table:n5128sim}
  \begin{center}
    \begin{tabular}{lc}
      \hline\hline
      {Parameter} & {Value} \\
      \hline
Number of Clusters & \\
Blue & 278 \\
Red & 310 \\
Radial Distribution & Modified Hubble Profile \\
Blue Population & \\
$\sigma_0$ & 0.2 arcmin$^{-2}$ \\
$R_0$ & 3.7' \\
Red Population & \\
$\sigma_0$ & 0.2 arcmin$^{-2}$ \\
$R_0$ & 4.0' \\
Angular Distribution & Spherically Symmetric \\
Luminosity Function & Gaussian \\
$\langle M_V \rangle$ & -8.0 \\
$\sigma_{M_V}$ & 1.0 \\
Velocity Dispersion & \citet{woodley10b} \\
Blue Population & \\
$\langle v \rangle$ & 26.0 km/s \\
$\sigma_v$ & 149.0 km/s \\
Red Population & \\
$\langle v \rangle$ & 43.0 km/s \\
$\sigma_v$ & 156.0 km/s \\

      \hline\hline
    \end{tabular}
  \end{center}
\end{table}

\section{Results}\label{sfour}

\subsection{The Isotropic and Tidally Filling Case}\label{sect:isotropic}

\begin{figure}
\centering
\includegraphics[width=\columnwidth]{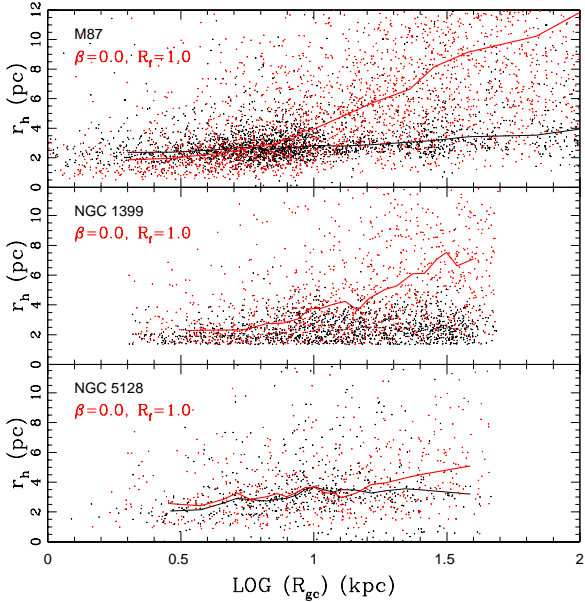}
\caption{ $r_h$ vs log $R_{gc}$ for observed globular clusters (black) and model clusters (red) in M87 (Top), NGC 1399 (Middle), and NGC 5128 (Bottom) assuming clusters have an isotropic distribution of orbits and are all tidally filling. The solid lines represent the median $r_h$ calculated with radial bins containing $5\%$ of the observed cluster population.}
  \label{fig:rh_B0}
\end{figure}

We first compare our observed datasets to a baseline set of models in which the clusters are tidally filling with an isotropic distribution of orbits. To best compare to observations the three dimensional positions of our model clusters are projected onto a two-dimensional plane. In Figure \ref{fig:rh_B0} we have plotted the measured $r_h$ of observed GCs (black) and theoretically determined $r_h$ of model clusters (red) in all three galaxies. The solid lines show the median $r_h$ as a function of $R_{gc}$.

Given the radial distribution of GCs and the mass profiles of M87 and NGC 1399, assuming each model cluster has a circular orbit at its current $r_{gc}$ would yield values of $\alpha$ equal to 0.6. For NGC 5128, despite its shallower mass profile the radial distribution of GCs is such that $\alpha=0.46$ (again assuming all clusters have circular orbits). Unfortunately, especially in the cases of M87 and NGC 1399, an isotropic distribution of orbits does not eliminate the difference between observed and theoretically predicted vales of $\alpha$. For M87 and NGC 1399, allowing clusters to have an isotropic distribution of orbits results in $\alpha = 0.55$. In the case of NGC 5128, the observations are surprisingly well matched by the isotropic case with $\alpha = 0.25$, with the model only slightly over-estimating cluster sizes at large $R_{gc}$. NGC 5128 is better fitted by the $\beta=0$ and $R_f=1$ model than M87 and NGC 1399 because its mass profile and smaller observed velocity dispersion results in clusters having a higher mean eccentricity and a smaller $\alpha$. Since clusters are brought deeper into the potential well of the galaxy, they must also be more tidally filling (despite the galaxy being less massive) to match the higher global value of $\alpha = 0.19$ observed in NGC 5128.

Kinematic and structural studies of M87 and NGC 1399 do not support the idea of cluster populations being isotropic and tidally filling.  Studies of galaxy formation and structure suggest that cluster orbits become preferentially radial and clusters become more tidally under-filling with increasing $r_{gc}$. As previously discussed in this study and in \citet{webb13a}, $\alpha$ will be further decreased by allowing either the anisotropy parameter $\beta$ to increase or the tidal filling parameter $R_f$ to decrease. Increasing $\beta$ serves to decrease cluster sizes as it results in cluster orbits being preferentially radial, bringing them deeper into the galactic potential of the galaxy. Decreasing $R_{f}$ also results in clusters being compact and tidally under-filling, such that their observed size is less than $r_t$ and they evolve as if they were in isolation. However, in \citet{webb13a} we only studied the effects of radially constant values of $\beta$ and $R_f$ on GC sizes. We explore the effects of these two parameters changing with  $r_{gc}$ in the following sub-sections.

\subsection{The Effects of Orbital Anisotropy and Tidally under-filling Clusters}\label{sect:global}

\begin{figure*} 
\centering
\includegraphics[width=\textwidth]{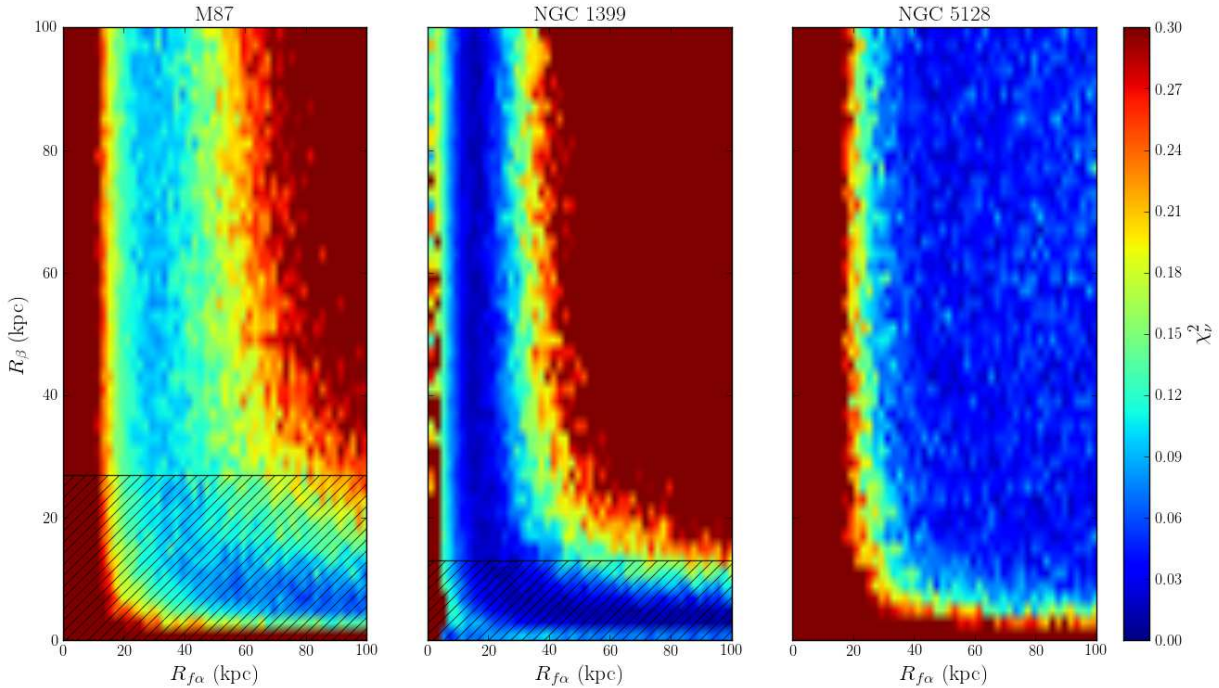}
\caption{Degeneracy between $R_\beta$ and $R_{f\alpha}$ for fits to the total cluster populations of M87 (left), NGC 1399 (middle) and NGC 5128 (right). Orbits go from preferentially radial to isotropic as $R_\beta$ increases and clusters become preferentially tidally filling as $R_{f\alpha}$ increases. The colour scale corresponds to the $\chi^2_{\nu}$ between our theoretical model and observations. Hatched out regions can be excluded based on kinematic studies of each galaxy (see Sections 4.4.1-4.4.3). } 
  \label{fig:degen}
\end{figure*}

To explore the effects of radially dependent $\beta$ and $R_f$, we re-run our simulations for $0 < R_\beta < 100 $ kpc and $0 < R_{f\alpha} < 100$ kpc in search for the combination which provides the strongest agreement between our theoretical and observed cluster populations. We have initially assumed that red and blue clusters in each galaxy have the same orbital anisotropy and tidal filling profiles. To compare theory and observations, we determine the median $r_h$ in 20 radial bins each containing $5\%$ of the total population. A median half-light radius is also calculated for each mock GC population using the same radial bins as the observations. To measure how well a model reproduces the observed dataset, we calculate the $\chi^2_{\nu}$ between the two median profiles via:

\begin{equation} \label{chi2}
\chi^2_{\nu} =\frac{1}{N-n-1} \sum\limits^N_{i} \frac{ (r_{h,obs}(R_{gc,i})-r_{h,mod}(R_{gc,i}))^2}{r_{h,obs}(R_{gc,i})+r_{h,mod}(R_{gc,i})}
\end{equation}

\noindent where $N$ is the total number of bins (20), n is the total number of fitted parameters (2), $r_{h,mod}(R_{gc,i})$ is the median half-light radius of the model in the $i^{th}$ radial bin and $r_{h,obs}(R_{gc})$ is the median half-light radius of the observations in the $i^{th}$ radial bin. The $\chi^2_{\nu}$ value between our model and the observations is shown for the entire $R_\beta$ and $R_{f\alpha}$ parameter space in Figure \ref{fig:degen}.

For M87 and NGC 1399, we see that high $R_\beta$ - low $R_{f\alpha}$ models can fit the observations just as well as low $R_\beta$ - high $R_{f\alpha}$ models. The degeneracy is due to the previously mentioned fact that both parameters are used to decrease cluster sizes. In NGC 5128 the degeneracy is less clearly defined due to the smaller number of clusters and their significantly smaller observed velocity dispersion. A smaller velocity dispersion means that even moderate changes in $R_\beta$ will not strongly affect the global kinematic properties of the model cluster population. With a lower number of clusters to sample the velocity dispersion with, the effect that changing $R_\beta$ has on the model population is minimized further. Only very low values of $R_\beta$, such that the relative values of $\sigma_r$, $\sigma_\theta$, and $\sigma_\phi$ differ dramatically, will the kinematic and structural properties of model clusters be noticeably altered. Hence any combination of moderate to large values of both $R_\beta$ and $R_{f\alpha}$ yield a model population that is primarily isotropic and provides a match between theoretical and observed cluster sizes.

A second key issue that adds to the degeneracy between $R_\beta$ and $R_{f\alpha}$ is the limited range in $R_{gc}$ of our datasets. It is difficult to rule out higher values of $R_\beta$ or $R_{f\alpha}$ based on cluster size alone since the majority of our observed clusters are found within 10 kpc of the galaxy centre. Hence high values of $R_\beta$ or $R_{f\alpha}$ will not change the anisotropy or tidal filling profiles of inner region clusters, and only affect the outermost clusters in each galaxy. In order to remove some of the degenerate solutions in Figure \ref{fig:degen} and identify acceptable values of $R_\beta$ and $R_{f\alpha}$ we must therefore look beyond our own observational datasets and models.

\subsection{Removing Degenerate Solutions} \label{degen}

The degeneracy between $R_\beta$ and $R_{f\alpha}$ in each galaxy indicates that there are multiple models which yield low values of $\chi^2_{\nu}$ between the observed and theoretical median $r_h$ profiles. To better constrain the parameter space, and eliminate some of the degeneracy between the two parameters, we can draw upon previous observational studies of each GC population. Kinematic studies of each galaxy have led to estimates of the global value of $\beta$, with some studies even suggesting possible anisotropy profiles. In order to eliminate some of our degenerate model solutions without placing undue weight on these previous studies, we only eliminate values of $R_\beta$ that produce cluster populations with mean values of $\beta$ that are outside the range of measured global values of $\beta$ or outside the $\beta$ range implied by an anisotropy profile. For a model consisting of $N$ GCs, the mean $\beta$ is simply $<\beta>=\frac{\sum_{n=1}^{N} \beta(n)}{N}$ where $\beta(n)$ is the value of $\beta$ at the given cluster's $r_{gc}$. $R_{f\alpha}$ on the other hand cannot as easily be constrained, as many issues including the initial size distribution of GCs, their tidal histories, and their merger/accretion histories can produce a range of different $R_f$ profiles. Therefore we will focus on finding values of $R_\beta$ that are also in agreement with kinematic studies of each galaxy.

\subsubsection{M87}

For M87, many kinematic studies exist that yield conflicting values of $\beta$ and $R_\beta$. A study by \citet{strader11} inferred high global values of $\beta$ ($\sim 0.4$). Studies by \citet{romanowsky01} and \citet{murphy11} on the other hand found that GCs in M87 are for the most part isotropic, with \citet{murphy11} inferring a small degree of radial anisotropy beyond 30 kpc. These three studies however go against the general findings of \citet{deason12}, who in a study of 15 elliptical galaxies concluded that the distribution of cluster orbits in elliptical galaxies are primarily isotropic with a slight preference towards tangential orbits in some cases. This result was also found in M87 by \citet{agnello14}. The most recent study regarding the kinematics of M87 by \citet{zhu14}, which has the largest GC kinematic dataset to date out to 180 kpc, suggested that inner cluster orbits are tangentially biased ($\beta=-0.2$) with $\beta$ increasing to 0.2 at 40 kpc and then decreasing back to 0 at 120 kpc. While the anisotropy profile suggested by \citet{zhu14} is in disagreement with \citet{strader11}, it agrees with the results of \citet{romanowsky01} and \citet{murphy11} while supporting the work of \citet{deason12} that the profiles might be partially tangentially biased. The fact that \citet{zhu14} found that $\beta$ begins to decrease again at larger $R_{gc}$ agrees with the general behaviour of $\beta$ predicted by \cite{agnello14}, but it does not support the idea that outer clusters have preferentially tangential orbits. Due to the large dataset and detailed method for determining the anisotropy profile of M87, we will use the results of \citet{zhu14} to remove degenerate model solutions that reproduce the distribution of cluster sizes in M87.

Based on the anisotropy profile suggested by \citet{zhu14}, we exclude degenerate solutions where the mean $\beta$ is greater than 0.2. Hence we can eliminate all models with $R_\beta$ less than 27 kpc. Based on this constraint and our model fits to the distribution of cluster sizes in M87, it appears that $R_{f\alpha}$ must be between 25 and 40 kpc. The narrow acceptable range in $R_{f\alpha}$ suggests that $R_{f\alpha}$ might be the dominant parameter in determining the size profile of GCs in M87, with $R_{f\alpha}$ marking the distance where clusters transition from being tidally affected to being tidally unaffected. Having $R_{f\alpha}$ within this range results in the outermost clusters having $R_f$ values between 0.06 and 0.14. Both of these values are comparable to the minimum $R_f$ values in the Milky Way, which are approximately equal to 0.1. And since the outer clusters in M87 (many of which could have been tidally truncated before being accreted by M87) orbit in weaker tidal fields than the majority of Galactic GCs, we do not feel that $R_{f\alpha}$ can be constrained any further without additional $r_h$ measurements of clusters beyond 40 kpc. To visualize the constraints that have now been placed on $R_\beta$ and $R_{f\alpha}$, models that do not agree with the results of \citet{zhu14} have been hatched out in Figure \ref{fig:degen}.

\subsubsection{NGC 1399}

For NGC 1399, the degeneracy is larger than in M87 because the observational dataset contains almost half as many GCs and spans only $\frac{1}{3}$ the range in $R_{gc}$. \citet{schuberth10} modelled the cluster populations with $\beta$ values between 0 and 0.5. Assuming the mean global $\beta$ is less than 0.5, then $R_\beta$ must be greater than 13 kpc. Similarily to M87, this constraint and our model fits to the distribution of cluster sizes allows us to set $R_{f\alpha}$ between 5 and 35 kpc. Since our NGC 1399 dataset only reaches out to 40 kpc, $R_{f\alpha}$ cannot be constrained any further and we can only conclude that the high $R_\beta$ - low $R_{f\alpha}$ region (upper left of Figure \ref{fig:degen}) is acceptable. A detailed studied of GC sizes in the outer regions of NGC 1399 will likely allow us to constrain $R_f$ even further. For visualization purposes, models outside of the kinematically constrained range have been hatched out of Figure \ref{fig:degen}.

Comparing the acceptable ranges in $R_{f\alpha}$ for M87 and NGC 1399 indicates that $R_{f\alpha}$ is likely a reflection of tidal field strength. If we refer back to Figure \ref{fig:mass}, $M_g(r_{gc})$ increases at a slower rate in NGC 1399 resulting in $r_t$ increasing at a faster rate compared to M87. If we assume the initial distribution of cluster sizes and their subsequent expansion is self-similar between galaxies, then the $r_{gc}$ at which GCs are no longer tidally affected will be smaller in NGC 1399. Hence the allowed range in $R_{f\alpha}$ for NGC 1399 should centre around a smaller $R_{gc}$ than M87, as observed. Furthermore, since the range in $R_{f\alpha}$ suggests that clusters in both galaxies become under-filling rather quickly, it is possible that the mild increase in $r_h$ with $R_{gc}$ observed in these galaxies is imprinted upon cluster formation. Hence only the innermost GCs and GCs with highly eccentric orbits will have their structural parameters altered by the tidal field of the galaxy.

\subsubsection{NGC 5128}

In a kinematic study of NGC 5128, \citet{woodley10b} also found that clusters could be approximated as having an isotropic distribution of orbits, with only a minor degree of radial anisotropy in the outermost regions (if at all). This is not surprising, since NGC 5128 was  well fitted by an isotropic and tidally filling GC population in Section \ref{sect:isotropic}. Unfortunately, since no upper limit was placed on the global value of $\beta$ we cannot constrain the parameter space beyond the $\chi^2_{\nu}$ values presented in Figure \ref{fig:degen}. Hence we are forced to consider all models with $R_\beta$ greater than 5 kpc and $R_{f\alpha}$ greater than 30 kpc. So while our model and kinematic studies support the conclusion that NGC 5128 is most likely isotropic and tidally-filling, we cannot rule out a slow increase in $\beta$ or decrease in $R_f$ with $r_{gc}$ based on the kinematic and structural studies presented here.

\subsection{The Orbital Anisotropy and Tidal Filling Profiles of M87, NGC 1399, and NGC 5128}\label{rvsb}

\begin{figure}
\centering
\includegraphics[width=\columnwidth]{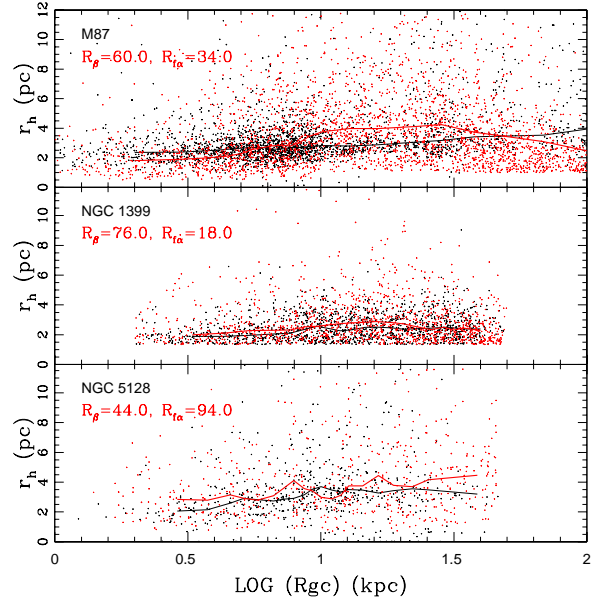}
\caption{ $r_h$ vs log $R_{gc}$ for observed globular clusters (black) and model clusters (red) in M87 (Top), NGC 1399 (Middle), and NGC 5128 (Bottom). Model clusters have anisotropy and tidal filling profiles as given by Equations \ref{beta_prof} and \ref{rf_prof}, with the best fit values of $R_\beta$ and $R_{f\alpha}$ indicated in each panel. The solid lines represent the median $r_h$ calculated with radial bins containing $5\%$ of the observed cluster population.}
  \label{fig:rh_prof}
\end{figure}

\begin{figure}
\centering
\includegraphics[width=\columnwidth]{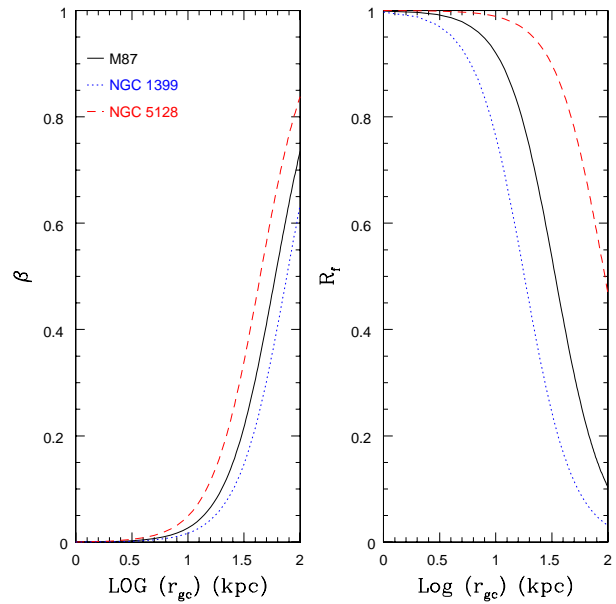}
\caption{The anisotropy profile (left panel) and filling profile (right panel) that produce the theoretical distribution of cluster sizes that best matches the observed distributions in M87 (black), NGC 1399 (blue) and NGC 5128 (red).}
  \label{fig:profs}
\end{figure}

We are now in a position to use our models to estimate the true values of $R_\beta$ and $R_{f\alpha}$. The best fit models are illustrated for each galaxy in Figure \ref{fig:rh_prof}. In all cases the $\chi^2_{\nu}$ between the observed and theoretical median $r_h$ profiles is less than unity. We also have plotted in Figure \ref{fig:profs} the $\beta(r_{gc})$ and $R_f(r_{gc})$ profiles which correspond to each best fit model.


The allowed range of models for M87 has $R_\beta > 27$ kpc and $20 < R_{f\alpha} < 40$ kpc. NGC 1399 on the other hand has a wider range of acceptable models with $R_\beta > 13$ kpc and $10 < R_{f\alpha} < 30$ kpc. However within these ranges, the best fit models to M87 and NGC 1399 are very similar, with M87 having a $R_\beta = 60$ kpc and $R_{f\alpha} = 34$ kpc and NGC 1399 having $R_\beta = 76$ kpc and $R_{f\alpha} = 18$ kpc. So while cluster orbits become more radial with $r_{gc}$, it appears that the decrease in $R_f$ is relatively steep, reaching values of 0.5 at 34 kpc and 18 kpc for M87 and NGC 1399 respectively. At the same time, the orbital anisotropy parameter $\beta$ reaches 0.5 at approximately 60 kpc and 76 kpc in M87 and NGC 1399 respectively. Hence our models suggest NGC 1399 might be slightly more isotropic than M87, in agreement with \citet{strader11}. 

The accepted range of models for NGC 5128 is quite different from either M87 or NGC 1399, with $R_\beta > 5$ kpc and $R_{f\alpha} > 30$ kpc. The best fit model was $R_\beta = 44$ kpc and $R_{f\alpha} = 94$ kpc. However, as previously mentioned, the lower velocity dispersion and small number of clusters in NGC 5128 means that even with $R_\beta = 44$ kpc the model population is not significantly different from the $\beta=0$ case. The high $R_{f\alpha}$ is required because the mean eccentricity is higher in NGC 5128 than the other two galaxies (given a mean $\beta \sim 0$) and clusters must be tidally filling to reproduce the higher observed $\alpha$. It should be noted though that there is still a large degree of uncertainty associated with the best fit model to NGC 5128 due to the significant amount of degeneracy between $R_\beta$ and $R_{f\alpha}$. In fact, we cannot clearly distinguish between the best fit model in Figure \ref{fig:profs} and other model solutions that also accurately reproduce the distribution of cluster sizes. We even caution against introducing radial profiles in either $\beta$ or $R_f$ in the first place, since the $\chi^2_{\nu}$ for the NGC 5128 model with $\beta=0$ and $R_f=1$ is only slightly improved compared to invoking $\beta$ and $R_f$ profiles.

Comparing Figure \ref{fig:rh_prof} to the isotropic and tidally filling cases (Figure \ref{fig:rh_B0}), we see that for M87 and NGC 1399 we have significantly improved the discrepancy between theoretical and observed cluster sizes. However, in the case of M87, the discrepancy is still not completely removed. In M87, our model slightly underestimates cluster sizes within 10 kpc, over estimates cluster sizes between 10 and 50 kpc, and again underestimates cluster sizes in the outer regions. To achieve a better agreement between theoretical and observational sizes, we take a closer look in the next section at the properties of the red and blue sub-populations in each galaxy.

\subsection{Separating the Metal Rich and Metal Poor Sub-Populations}

In the previous section, we made the initial assumption that all clusters in a single galaxy share the same $\beta$ and $R_f$ profiles. However, it has long been known that GC populations in many types of galaxies can be divided into at least two sub-populations based on colour \citep[e.g.][] {zepf93, larsen01, harris09b, peng06}. Colour bimodality within cluster populations is often attributed to metallicity, with metal poor clusters being bluer than metal rich clusters \citep[e.g.][] {zepf93, brodie06}. Since this bimodality is observed over a wide range of galaxy masses and types \citep{harris15}, it is believed that the production of a two (or more) component GC population is an important step inherent to all galaxy formation and evolution mechanisms.

\begin{figure}
\centering
\includegraphics[width=\columnwidth]{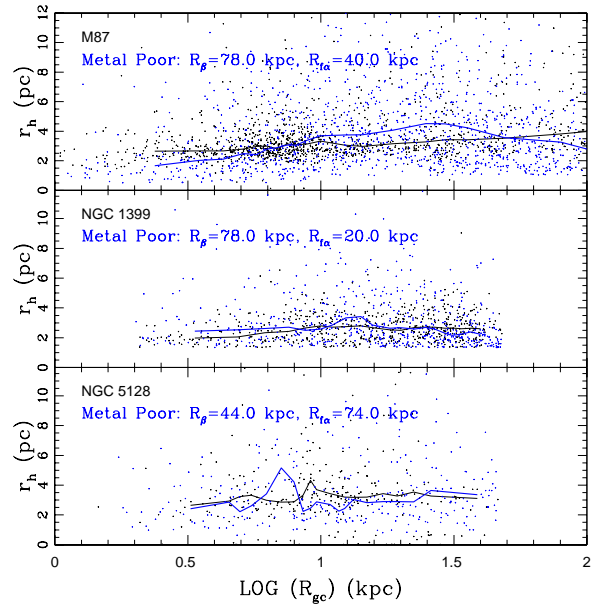}
\caption{ $r_h$ vs log $R_{gc}$ for observed metal poor globular clusters (black) and model clusters (blue) in M87 (Top), NGC 1399 (Middle), and NGC 5128 (Bottom). Model clusters have anisotropy and tidal filling profiles as given by Equations \ref{beta_prof} and \ref{rf_prof}, with the best fit values of $R_\beta$ and $R_{f\alpha}$ indicated in each panel. The solid lines represent the median $r_h$ calculated with radial bins containing $5\%$ of the observed cluster population.}
  \label{fig:rh_blue}
\end{figure}

\begin{figure}
\centering
\includegraphics[width=\columnwidth]{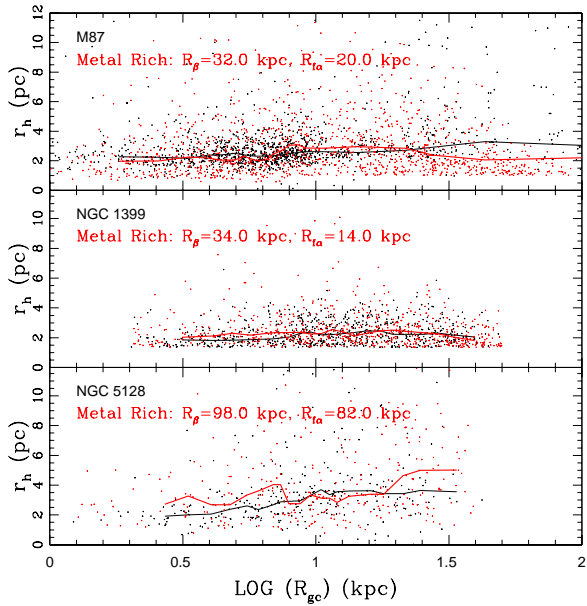}
\caption{ $r_h$ vs log $R_{gc}$ for observed metal rich globular clusters (black) and model clusters (red) in M87 (Top), NGC 1399 (Middle), and NGC 5128 (Bottom). Model clusters have anisotropy and tidal filling profiles as given by Equations \ref{beta_prof} and \ref{rf_prof}, with the best fit values of $R_\beta$ and $R_{f\alpha}$ indicated in each panel. The solid lines represent the median $r_h$ calculated with radial bins containing $5\%$ of the observed cluster population.}
  \label{fig:rh_red}
\end{figure}

Observational studies have identified many structural and kinematic differences between these two sub-populations. A common observation within GC populations (including the populations presented here) is that red GCs have half-light radii that are on average $20 \%$ ($\sim 0.4$ pc) smaller than blue GCs \citep[e.g.][]{kundu98, kundu99, larsen01, jordan05, harris09a, harris10a, paolillo11, blom12, strader12, woodley12, usher13}. The size difference is likely due to the red and blue sub-populations having different formation, dynamical and stellar evolution histories \citep[e.g.][]{kundu98, jordan04, jordan05, harris09a, sippel12, schulman12}. As discussed in Section \ref{degen}, kinematic studies of GC populations find that red and blue sub-populations have noticeably different radial profiles and velocity dispersions, with some studies even suggesting that red and blue clusters have different $\beta$ profiles as well \citep[e.g][]{cote01,schuberth10}. Therefore, we have elected to repeat the fitting process, but with the red and blue clusters modelled separately. The final comparison between our models and observations is illustrated in Figure \ref{fig:rh_blue} for metal poor clusters and Figure \ref{fig:rh_red} for metal rich clusters.

By modelling metal poor and metal rich clusters separately we slightly improve the $\chi^2_{\nu}$ between the observed and theoretical median $r_h$ profiles of M87 and NGC 1399, with all $\chi^2_{\nu}$ values again being less than 1. For M87, our model suggests that red clusters have a smaller $R_\beta$ (more radial orbits) than blue clusters. Our study also finds that red clusters become tidally under-filling very quickly with $r_{gc}$ compared to blue clusters. The different $R_f$ profiles are in agreement with observational studies that found red clusters are on average smaller than blue clusters at all $R_{gc}$. For NGC 1399, the best fit anisotropy profiles to red and blue clusters are almost identical to M87. Also similar to M87, red clusters in NGC 1399 have lower vales of $R_\beta$ and $R_{f\alpha}$ than blue clusters. The only difference between the two galaxies is that clusters in NGC 1399 become under-filling slightly quicker than M87, which is likely a result of the tidal field in NGC 1399 being weaker than in M87 since it is less massive.

In NGC 5128, allowing red and blue clusters to have different values of $R_\beta$ and $R_{f\alpha}$ yields much higher $\chi^2_{\nu}$ values than either the isotropic and tidally filling case or when the sub-populations were given the same $R_\beta$ and $R_{f\alpha}$ values. While this observation may be a result of the previously discussed issues regarding the degeneracy between $R_\beta$ and $R_{f\alpha}$ in NGC 5128, it appears the galaxy is still best described as a singular tidally filling population with a mostly isotropic distribution of orbits.

\section{Discussion} \label{sfive}

Our model reproduces a realistic GC system by allowing cluster orbits to become more radial with $r_{gc}$, letting clusters become less tidally filling with $r_{gc}$ and includes the possibility of modelling the red and blue cluster sub-populations separately. After applying the model to M87, NGC 1399, and NGC 5128 we can now discuss the best fit profiles to each galaxy in further detail.

\subsection{M87}

The lowest $\chi^2_{\nu}$ between our model and observed GCs occurred when red and blue clusters were modelled independently. The fact that the best fit model for red and blue clusters implies the population becomes radially anisotropic with $r_{gc}$ is in agreement with the idea that giant galaxies form through the hierarchical merging of dwarf galaxies that combine to form a central massive galaxy\citep{kravtsov05, tonini13, li14, kruijssen14}. While some clusters will form early in the small halos which make up the host galaxy, most of the GC population in a giant galaxy has been added via the accretion of dwarf galaxies. With \citet{wu14} finding that massive galaxies with a large population of accreted stars will have a high degree of radial anisotropy at large $r_{gc}$, a similar result can be expected for outer GCs in massive galaxies. The identification of shells, arcs, and streams of kinematically distinct GCs in M87 suggests that the outer regions of M87 have been built up via a continuous infall of material that is still ongoing today \citep{strader11, romanowsky12, dabrusco13, dabrusco14a, dabrusco14b, dabrusco15, longobardi15}. \citet{strader11} and \citet{romanowsky12} also suggested that clusters beyond 40 kpc are being dynamically perturbed by nearby galaxies. Continuous perturbations could also result in many of the outer clusters being energized to eccentric orbits and further increase the degree of radial anisotropy in the outer regions of brightest cluster galaxies (BCGs).

Models of giant galaxies that form via the accretion of smaller galaxies also found the majority of clusters which form in the central host are metal-rich while the majority of accreted clusters are metal poor, suggesting that metal poor clusters may have a higher degree of orbital anisotropy than metal rich ones \citep{kruijssen15}. Our models do not support this statement, suggesting that kinematic evidence of cluster accretion may become non-existent not long after a dwarf merger event. Minor mergers with galaxies that contain their own bi-modal GC population may also erase any relationship between orbital anisotropy and cluster type that was established when the central galaxy first formed. However, we also cannot rule out greater differences between the red and blue anisotropy profiles due to the clear degeneracy between $R_\beta$ and $R_{f\alpha}$. We do find evidence in M87 that metal rich clusters are more under-filling than blue clusters at all $R_{gc}$, which indicates metal rich clusters either form more compact than metal poor clusters or expand at a slower rate.

Unfortunately, our best fit model to M87 still produces too many small clusters at low and high $R_{gc}$. Factors which may explain this difference between our model and observed cluster sizes include our choice for the functional form of the $\beta$ profile in Equation \ref{beta_prof}, our assumption that $n(R_{gc})$ continues beyond the observational dataset and the possible existence of a third cluster sub-population with an intermediate metallicity between red and blue clusters found by \citet{strader11} and \citet{agnello14}. With respect to the functional form of the $\beta$ profile, allowing inner clusters to have preferentially tangential orbits and outer clusters to have a near isotropic distribution of orbits (the latter of which was found by \citet{zhu14}) would increase the mean $r_h$ of clusters in the inner and outer regions of M87. Additionally, in the \citet{agnello14} model of M87, the velocity dispersion for each sub-population changes as a function of $R_{gc}$ which would also affect how the distribution of cluster orbits changes with $R_{gc}$. Future applications of our model will take into consideration the existence of more than two sub-populations, radially dependent velocity dispersions, different extrapolations of $n(R_{gc})$ and different functional forms of $\beta(r_{gc})$.

\subsection{NGC 1399}

The model which best reproduces the observed distribution of cluster sizes in NGC 1399 is also found when modelling the red and blue sub-populations separately. The best fit values of $R_\beta$ and $R_{f\alpha}$ in NGC 1399 are quite similar to M87. The only difference between the two populations is the clusters in NGC 1399 appear to be a bit more under-filling than clusters in M87, which is consistent with the tidal field of NGC 1399 being weaker than M87. Unfortunately the degeneracy between the two parameters in each galaxy are such that we cannot conclude whether or not both galaxies have the \textit{exact} same anisotropy and tidal filling profiles. It is interesting to note that our model of NGC 1399 predicts the same flattening in the $r_h$-$R_{gc}$ profile at large distances that our M87 model did, but in the case of NGC 1399 the observational profile supports this trend. Hence the decrease in $\beta$ inferred by \citet{agnello14} and \citet{zhu14} either doesn't occur in NGC 1399 and our functional form of $\beta(r_{gc}$) is correct or the cluster population has not be studied out to large enough $R_{gc}$. 

The general case of orbits becoming more radial and clusters becoming more under-filling with $r_{gc}$ is still prevalent in both M87 and NGC 1399. The fact that both galaxies are at the centre of large clusters is likely the common factor, as they formed via the hierarchical merging of smaller galaxies and neighbouring galaxies are continuously perturbing the outer cluster populations. Hence the outer regions of these galaxies cannot truly reach equilibrium. As we will see for NGC 5128 in the following section, a galaxy in isolation may in fact come closer to some sort of dynamical equilibrium.

\subsection{NGC 5128}

The mass profile and radial distribution of GCs in NGC 5128 is such that $\alpha$ is already quite low (0.46) if cluster orbits are all assumed circular. Allowing for an isotropic distribution of orbits, with the smaller velocity dispersion of NGC 5128, results in clusters having a higher mean eccentricity than either M87 or NGC 1399 and therefore a smaller mean size at a given $r_{gc}$. Furthermore, assuming an isotropic distribution of orbits, the radial distribution and velocity dispersion of GCs in NGC 5128 combine with its mass profile to yield a shallower increase in $r_h$ with $R_{gc}$ than either M87 or NGC 1399. Hence for NGC 5128, the distribution of cluster sizes is already well reproduced assuming the red and blue cluster populations are both primarily isotropic and tidally filling. For models where $\beta$ and $R_f$ are functions of $r_{gc}$, we see the degeneracy between $R_\beta$ and $R_{f\alpha}$ in NGC 5128 is quite different from either M87 or NGC 1399. While differences between the galaxies can be attributed to them having different mass profiles and GC radial distributions, the key issues are the lower velocity dispersion and small number of the observed GCs in NGC 5128 compared to M87 and NGC 1399.

The lower velocity dispersion and smaller number of the observed GCs in NGC 5128 result in there being almost no significant difference between the kinematic and structural properties of models with $\beta=0$ and $R_f=1$ and models with moderate to high values of $R_\beta$ and $R_{f\alpha}$. And as discussed in Section \ref{rvsb}, we were unable to constrain the degeneracy between $R_\beta$ and $R_{f\alpha}$ based on kinematic studies of NGC 5128. Trying to model the red and blue cluster populations separately also did not improve the fit to the observations, indicating both sub-populations have the same kinematic properties. With \citet{woodley10b} suggesting that the kinematic properties of NGC 5128 indicate very little (if any) radial anisotropy is present in NGC 5128, any model with $<\beta> \sim 0$ cannot be ruled out. Measuring GC sizes over a larger range in $R_{gc}$ and a more detailed study of anisotropy in NGC 5128 are necessary in order to place stronger constraints on $R_\beta$ and $R_{f\alpha}$ and minimize the uncertainty associated with best fit anisotropy and tidal filling profiles.

\textit{If} NGC 5128 has a primarily isotropic population, it would suggest that NGC 5128 was assembled during an initial fast accretion phase \citep[e.g.][]{biviano09} and has undergone few recent major mergers. Hence over 12 Gyr, any clusters accreted during this initial fast accretion will have their radial orbits decay and will have been pulled towards the centre of the galaxy \citep{goodman84, lee89, cipolina94}. NGC 5128 is known to have several observational features (e.g. central black holes, jets, dust lanes) that are also seen in the majority of giant elliptical galaxies and are consistent with a history of mergers and on-going accretion events \citep{vandokkum03,harris10b, rejkuba11}. However, since NGC 5128 is not a BCG like M87 and NGC 1399, it has likely not undergone as many mergers or accretion events. Furthermore, NGC 5128 does not have any massive galaxies or satellites nearby to perturb the outer cluster population. 

\section{Conclusions and Future Work} \label{ssix}

We have successfully reproduced the distributions of GC sizes in three giant galaxies (M87, NGC 1399, and NGC 5128) by allowing cluster orbits to become more radial and clusters to become more under-filling with $r_{gc}$, in line with models and observations of galaxy structure and cluster populations \citep[e.g.][]{cote01,prieto08, zait08, weijmans09, gnedin09,ludlow10, strader12, kruijssen12,alexander13, puzia14}. For M87 and NGC 1399, both galaxies that are located at the centres of galaxy clusters, the global cluster populations have a high degree of radial anisotropy at larger $r_{gc}$ and are primarily under-filling in the outer regions. Our findings are consistent with kinematic studies of each galaxy \citep[e.g][]{cote01,schuberth10, woodley10b, woodley10c,murphy11,agnello14, zhu14}, the assembly of giant galaxies via mergers and dwarf galaxy accretion \citep[e.g][]{schuberth10, kruijssen12}, and the evolution of GC populations \citep{alexander13, webb13a}. NGC 5128 on the other hand was more difficult to model due to the lower number of observed clusters, but to first order appears to be nearly isotropic and tidally filling out to large $R_{gc}$.

The best fit orbital anisotropy and filling profiles of each of these galaxies come with significant uncertainty due to the strong degeneracy between $R_\beta$ and $R_{f\alpha}$. Both parameters serve to decrease cluster size with $r_{gc}$. For M87 and NGC 1399, both datasets can be fitted by either a low $R_\beta$ - high $R_{f\alpha}$ or low $R_{f\alpha}$ - high $R_\beta$ model. However, kinematic studies of M87 and NGC 1399 allow us to rule out the low $R_\beta$ - high $R_{f\alpha}$ cases, and we can accurately interpret M87 as having $R_\beta > 27$ kpc and $20 < R_{f\alpha} < 40$ kpc and NGC 1399 having $R_\beta > 13$ kpc and $10 < R_{f\alpha} < 30$ kpc. The best fit models are $R_\beta = 60$ kpc and $R_{f\alpha} = 34$ for M87 and $R_\beta = 76$ kpc and $R_{f\alpha} = 18$ for NGC 1399. Hence in both galaxies the orbits become moderately radial and clusters become tidally under-filling with $r_{gc}$. The fact that the acceptable range in $R_{f\alpha}$ is lower for NGC 1399 is consistent with the galaxy's mass profile increasing at a shallower rate, which suggests the present day relationship between $r_h$ and $R_{gc}$ is set upon cluster formation. Assuming clusters in both galaxies formed with the same initial distribution in $r_h$, only the innermost clusters and clusters with eccentric orbits will have expanded to the point of becoming tidally filling, with clusters in NGC 1399 becoming tidally under-filling at a lower $R_{gc}$ than M87.

Unfortunately, since NGC 5128 is best fitted by models with $R_\beta > 5$ kpc and $R_{f\alpha} > 30$ kpc the degeneracy between the two parameters is much different than in M87 and NGC 1399. We attribute the larger uncertainty in $R_\beta$ and $R_{f\alpha}$ to the lower velocity dispersion and fewer number of clusters in NGC 5128. So while kinematic studies of NGC 5128 suggest the population is primarily isotropic, we cannot confidently rule out models where $\beta$ and $R_f$ gradually change with $R_{gc}$.

Studying the models which best reproduce the distribution of red and blue cluster sizes in each galaxy individually, given the constraints implied by previous kinematic studies, also provides useful information regarding each galaxy. In both M87 and NGC 1399 we find that red clusters are significantly more under-filling than blue clusters at all $R_{gc}$, which suggests they formed more compact or expand at a slower rate. Furthermore, the best fit orbital anisotropy profiles of red and blue clusters in M87 and NGC 1399 are almost identical, suggesting the orbital anisotropy and tidally filling profiles of BCGs in general can be quite similar. If this is true, it would suggest that differences in the anisotropy profiles of accreted and non-accreted clusters will not exist if accretion took place a long time ago and the entire cluster population has had time to dynamically mix. Hence only the outermost clusters, which have yet to mix and are being perturbed by neighbouring satellites, will be found on radial orbits. Additional observations in each galaxy at large $R_{gc}$ would be necessary to properly compare metal-poor and metal-rich clusters that may have just recently joined their central host. 

For NGC 5128, modelling the red and blue GC populations separately actually results in a worse fit to observations than when the sub-populations are considered to have the same kinematic properties. This finding suggests the two sub-populations have had time to fully mix. Analysis of our models also suggests that NGC 5128 could be very different from either M87 or NGC 1399, as we infer that NGC 5128 has a nearly isotropic and tidally filling population of GCs. This difference may suggest NGC 5128 formed via a fast accretion phase. A fast-accretion phase, combined with the fact that NGC 5128 is an isolated galaxy with no major satellites to perturb the outer cluster population, could result in an isotropic distribution of cluster orbits at all $R_{gc}$. 

Detailed kinematic models of cluster sub-populations which take into consideration how velocity dispersion changes a function of $r_{gc}$ are required in order for us to place stronger constraints on $R_\beta$. Measuring cluster sizes and velocities over a larger range in $R_{gc}$ will also help minimize the degeneracy between $R_\beta$ and $R_{f\alpha}$. In order to better constrain $R_{f\alpha}$, models of how cluster mass and structure evolves in the tidal fields of giant galaxies are needed, including clusters that have been accreted via mergers. Ultimately, what we can say with more certainty is that best fit models to M87 and NGC 1399 are telling us that BCGs can have different formation and merger histories but still have their present-day GC populations evolving in similar environments. The best model fits to NGC 5128 on the other hand indicate that more isolated giant galaxies are different then their brightest cluster galaxies counterparts. Hence differences between forming via an initial fast accretion stage followed by either slow or minor accretions/mergers versus having a longer dynamical history and neighbouring perturbing galaxies all leave different imprints on the cluster population in the form of its orbital anisotropy and tidal filling profiles.

\section{Acknowledgements}

We would like to thank the referee for constructive comments and suggestions regarding the presentation of the paper. JW, WEH and AS also acknowledge financial support through research grants and scholarships from the Natural Sciences and Engineering Research Council of Canada. M.P. acknowledges financial support from PRIN-INAF 2014 ''Fornax Cluster Imaging and Spectroscopic Deep Survey".


\bsp

\label{lastpage}


\begin{thebibliography}{99}

\bibitem[{{Agnello} {et~al.}(2014)}]{agnello14} Agnello, A., Evans, N.W., Romanowsky, A.J., Brodie, J.P. 2014, MNRAS, 442, 3299

\bibitem[{{Alexander} \& {Gieles}(2013)}]{alexander13} Alexander, P. E. R. \& Gieles, M. 2013, MNRAS, 432L, 1

\bibitem[\protect\citeauthoryear{Baumgardt \& Makino}{2003}]{baumgardt03} Baumgardt H., Makino J. 2003, MNRAS, 340, 227


 \bibitem[{{Bertin} \& {Varri}(2008)}]{bertin08} Bertin, G. \& Varri, A. L. 2008, ApJ, 689, 1005



\bibitem[{{Bianchini} {et~al.}(2015)}]{bianchini15} Bianchini, P., Renaud, F., Gieles, M., Varri, A.L. 2015, MNRAS, 447, 40

\bibitem[{{Binney} \& {Tremaine}(2008)}]{binney08} {Binney}, J. \& {Tremaine}, S. 2008, {Galactic dynamics second edition} (Princeton, NJ, Princeton University Press, 1987, 747 p.)

 \bibitem[{{Biviano} \& {Poggianti}(2009)}]{biviano09} Biviano, A. \& Poggianti, B. M. 2009, A\&A, 501, 419


\bibitem[{{Blakeslee} {et~al.}(2009)}]{blakeslee09} Blakeslee, J. P., J\'{o}rdan, A., Mei, S., C\^{o}t\'{e}, P., Ferrarese, L., Infante, L., Peng, E.W., Tonry, J.L., West, M.J. 2009, ApJ, 694, 556

\bibitem[{{Blom} {et~al.}(2012)}]{blom12} Blom, C., Spitler, L. R., Forbes, D. 2012, MNRAS, 420, 37

 \bibitem[{{Brodie} \& {Strader}(2006)}]{brodie06} Brodie, J. P. \& Strader, J. 2006, ARAA, 44, 193





 \bibitem[{{Casetti-Dinescu} {et~al.}(2007)}]{dinescu07} Casetti-Dinescu, D.I., Girard, T.M., Herrera, D., van Altena, W.E., L\'{o}pez, C.E., Castillo, D.J. 2007, AJ, 134, 195

\bibitem[{{Casetti-Dinescu} {et~al.}(2013)}]{dinescu13} Casetti-Dinescu, D.I., Girard, T.M., J\'{i}kov\'{a}, L., van Altena, W.F., Podest\'{a}, F., L\'{o}pez, C.E. 2013, AJ, 146, 33



\bibitem[{{Cipolina} \& {Bertin}(1994)}]{cipolina94} {Cipolina}, M. \& {Bertin}, G. 1994, AA, 288, 43

\bibitem[{{C\^{o}t\'{e}} {et~al.}(2001)}]{cote01} C\^{o}t\'{e}, P., McLaughlin, D.E., Hanes, D.A., Bridges, T.J., Geisler, D., Merrid, D., Hesser, J.E., Harris, G.L.H., Lee, M.G., 2001, ApJ, 559, 828, 257B

\bibitem[{{D'Abrusco} {et~al.}(2015)}]{dabrusco15} D'Abrusco, R., Fabbiano, G., Zezas, A. 2015, ApJ, 805, 26

\bibitem[{{D'Abrusco} {et~al.}(2014a)}]{dabrusco14a} D'Abrusco, R., Fabbiano, G., Brassington, N.J. 2014a, ApJ, 783, 19

\bibitem[{{D'Abrusco} {et~al.}(2014b)}]{dabrusco14b} D'Abrusco, R., Fabbiano, G., Mineo, S., Strader, J., Fragos, T., Kim, D.-W., Luo, B., Zezas, A. 2014b, ApJ, 783, 18

\bibitem[{{D'Abrusco} {et~al.}(2013)}]{dabrusco13} D'Abrusco, R., Fabbiano, G., Strader, J., Zezas, A., Mineo, S., Fragos, T., Bonfini, P., Luo, B., Kim, D.-W., King A 2013, ApJ, 773, 87

\bibitem[{{Deason} {et~al.}(2012)}]{deason12} Deason, A.J., Belokurov, V., Evans, N.W., McCarthy, I.G. 2012, ApJ, 748, 2


 \bibitem[{{de Vaucouleurs} \& {Nieto}(1978)}]{devaucouleurs78} de Vaucouleurs, G., \& Nieto, J.-L. 1978, ApJ, 220, 449

\bibitem[{Dehnen(1993)}]{dehnen93} Dehnen, W. 1993, MNRAS, 265, 250

\bibitem[{{Dinescu} {et~al.}(1999)}]{dinescu99} Dinescu, D.I., Girard, T.M., van Altena, W.E. 1999, AJ, 117, 1792


\bibitem[{{Dunn} \& {Jerjen}(2006)}]{dunn06} Dunn, L.P. \& Jerjen, H. 2006, AJ, 132, 1384

\bibitem[{{Fall} \& {Zhang}(2001)}]{fall01} Fall, S. M. \& Zhang, Q., 2001, ApJ, 561, 751





\bibitem[{{Gieles} {et~al.}(2011)}]{gieles11} Gieles, M., Heggie , D. C., Zhao H. 2011, MNRAS, 413, 2509


\bibitem[{{Gnedin} \& {Prieto}(2009)}]{gnedin09} Gnedin, O. Y., \& Prieto, J. L. 2009, in ESO Astrophysics Symp.: Globular
Clusters-Guides to Galaxies (Berlin: Springer), 323

\bibitem[{{G\'{o}mez} \& {Woodley}(2007)}]{gomez07} G\'{o}mez, M. \& Woodley, K.A. 2007, ApJ, 670, L105

\bibitem[{{Goodman} \& {Binney}(1984)}]{goodman84} {Goodman}, J. \& {Binney}, J.J. 1984, MNRAS, 207, 511


\bibitem[{Harris (1996) (2010 Edition)}]{harris96} Harris, W. E. 1996, AJ, 112, 1487, 2010 Edition

\bibitem[{Harris (2009a)}]{harris09a} Harris, W.E. 2009a, ApJ, 703, 939

\bibitem[{Harris (2009b)}]{harris09b} Harris, W.E. 2009b, ApJ, 699, 254

\bibitem[{Harris {et~al.}(2010)}]{harris10a} Harris, W.E., Spitler, L.R., Forbes, D.A., Bailin, J. 2010, MNRAS, 401, 1965

 \bibitem[{Harris (2010)}]{harris10b} Harris, G.L.H., Rejkuba, M., Harris, W.E. 2010, PASA, 27, 457

 \bibitem[{Harris {et~al.}(2015)}]{harris15} Harris, William E., Harris, G.L.H., Hudson M.J. 2015, ApJ, 806, 36
 
 \bibitem[\protect\citeauthoryear{Henon}{1961}]{henon61} Henon M. 1961, Annales d'Astrophysique, 24, 369
  
 \bibitem[{{Innanen}, {Harris}, \& {Webbink}(1983)}]{innanen83} Innanen, K. A., Harris, W.E., Webbink, R.F. 1983, AJ, 88, 338


\bibitem[{J\'{o}rdan(2004)}]{jordan04} J\'{o}rdan, A. 2004, ApJ, 613, L117

 \bibitem[{{J\'{o}rdan} {et~al.}(2005)}]{jordan05} J\'{o}rdan, A., C\^{o}t\'{e}, P., Blakeslee, J. P., Ferrarese, L., McLaughlin, D. E. , Mei, S., Peng, E. W., Tonry, J. L., Merrit, D., Milosavljevi\'{c}, M., Sarazin, C. L., Sivakoff, G. R., West, M. J., 2005, ApJ, 634, 1002


 \bibitem[{Kennedy (2014)}]{kennedy14} Kennedy, G.F. 2014, MNRAS, 444,3328

\bibitem[{King (1962)}]{king62} King, I. R. 1962, AJ, 67, 471



\bibitem[{{Kormendy} {et~al.}(2009)}]{kormendy09} Kormendy, J., Fisher, D.B., Cornell, M.E., Bender, R. 2009, ApJS, 182, 216

 \bibitem[{{Kravtsov} \& {Gnedin}(2005)}]{kravtsov05}  Kravtsov, A.V. \& Gnedin, O.Y. 2005, ApJ, 623, 650

\bibitem[{{Kruijssen} {et~al.}(2012)}]{kruijssen12} Kruijssen, J.M.D., Pelupessy, F.I., Lamers, H.J.G.L.M., Portegies Zwart, S.F., Bastian, N., Icke, V. 2012, MNRAS, 421, 1927

\bibitem[\protect\citeauthoryear{Kruijssen}{2014}]{kruijssen14} Kruijssen, J.M.D. 2014,  Classical and Quantum Gravity, 31, 244006

\bibitem[\protect\citeauthoryear{Kruijssen}{2015}]{kruijssen15} Kruijssen, J.M.D. 2015, MNRAS, 454, 1658

\bibitem[{{Kundu} \& {Whitmore}(1998)}]{kundu98} Kundu, A. \& Whitmore, B. C. 1998, AJ, 116, 2841

\bibitem[{{Kundu} {et~al.}(1999)}]{kundu99} Kundu, A., Whitmore, B. C., Sparks, W. B., Macchetto, F. D., Zepf, S. E., Ashman, K. M. 1999, ApJ, 513, 733

\bibitem[{{K\"{u}pper} {et~al.}(2010)}]{kupper10} K\"{u}pper, A. H. W, Kroupa, P, Baumgardt, H., Heggie , D. C. 2010, MNRAS, 407, 2260

 \bibitem[{{Larsen}(1999)}]{larsen99} Larsen, S. S. 1999, \aa, 139, 393

\bibitem[{{Larsen} {et~al.}(2001)}]{larsen01} Larsen, S. S., Brodie, J. P., Huchra, J. P., Forbes, D. A., Grillmair, C. J. 2001, AJ, 121, 2974


\bibitem[{{Lee} \& {Goodman}(1989)}]{lee89} {Lee}, M.H. \& {Goodman}, J. 1989, ApJ, 343, 594

  \bibitem[{{Li} \& {Gnedin}(2014)}]{li14}  Li, H. \& Gnedin, O.Y. 2014, ApJ, 796, 10

\bibitem[{{Longobardi} {et~al.}(2015)}]{longobardi15} Longobardi, A., Arnaboldi, M., Gerhard, O., Mihos, J. C. 2015, A\&A, 579, 135

\bibitem[{{Ludlow} {et~al.}(2010)}]{ludlow10} Ludlow, A. D., Navarro, J. F., Springler, V., Vogelsberger, M., Wang , J., White, S. D. M., Jenkins, A., \& Frenk, C. S. 2010, MNRAS, 406, 137


 \bibitem[{{Madrid} {et~al.}(2014)}]{madrid14} Madrid, J.P., Hurley, J.R., Martig, M., 2014, ApJ, 784, 95





\bibitem[{{McLaughlin} \& {van der Marel}(2005)}]{mclaughlin05} McLaughlin, D. E. \& van der Marel, R. P. 2005, ApJs, 161, 304

\bibitem[{McLaughlin(1999)}]{mclaughlin99} McLaughlin, D. E. 1999, ApJ, 512, L9

\bibitem[{{Miholics} {et~al.}(2014)}]{miholics14} Miholics, M., Webb, J., Sills, A., 2014, MNRAS, 445, 2872

\bibitem[{{Miholics} {et~al.}(2015)}]{miholics16} Miholics, M., Webb, J., Sills, A., 2016, MNRAS, 456, 240

\bibitem[{{Moreno} {et~al.}(2014)}]{moreno14} Moreno, E., Pichardo, B., Vel\'{a}zquez, H. 2014, ApJ, 793, 110

\bibitem[{{Murphy}, {Gebhardt} \& {Adams}(2011)}]{murphy11} Murphy, J.D., Gebhardt, K. \& Adams, J.J. 2011, ApJ, 729, 129

\bibitem[{{Navarro}, {Frenk} \& {White}(1997)}]{navarro97} Navarro, J. F., Frenk, C. S., \& White, S. D. M. 1997, ApJ, 490, 493




 \bibitem[{{Paolillo} {et~al.}(2002)}]{paolillo02} Paolillo, M., Fabbiano, G., Peres, G. Kim, D.-W. 2002, ApJ, 565, 883

\bibitem[{{Paolillo} {et~al.}(2011)}]{paolillo11} Paolillo, M., Puzia, T. H., Goudfrooij, P., Zepf, S. E., Maccarone, T. J., Kundu, A., Fabbiano, G., Angelini, L. 2011, ApJ, 736, 90 


\bibitem[{{Peng} {et~al.}(2004)}]{peng04a} Peng, E. W., Ford, H. C., \& Freeman, K. C. 2004, ApJ, 602, 685

\bibitem[{{Peng} {et~al.}(2004)}]{peng04b} Peng, E. W., Ford, H. C., \& Freeman, K. C. 2004, ApJ, 602, 705

\bibitem[{{Peng} {et~al.}(2006)}]{peng06} Peng, E.W., J\'{o}rdan, A., C\^{o}t\'{e}, P., Blakeslee, S., Ferrarese, L., Mei, S., West, M. J., Merritt, D., Milosavljevi\'{c}, M., Tonry, J. L. 2006, ApJ, 639, 95 

\bibitem[{{Peng} {et~al.}(2010)}]{peng10} Peng, E.W., Ho, L. C., Impey, C. D., \& Rix, H.-W. 2010, AJ, 139, 2097

 \bibitem[{{Pierce} {et~al.}(1994)}]{pierce94} Pierce, M. J., Welch, D. L., McClure, R. D., van den Bergh, S., Racine, R., \& Stetson, P. B. 1994,
Nature, 371, 385

\bibitem[{{Prieto} \& {Gnedin}(2008)}]{prieto08} Prieto, J. L. \& Gnedin, O. Y. 2008, ApJ, 689, 919 
 
 \bibitem[{{Puzia} {et~al.}(2014)}]{puzia14} Puzia, T.H., Paolillo, M., Goudfrooij, P., Maccarone, T.J., Fabbiano, G., Angelini, L. 2014, ApJ, 786, 78
 
 \bibitem[{{Rejkuba} {et~al.}(2011)}]{rejkuba11} Rejkuba, M., Harris, W.E., Greggio, L., Harris, G.LH. 2011, A\&A, 526, 123
  
 \bibitem[{{Renaud} {et~al.}(2011)}]{renaud11} Renaud, F., Gieles, M., Christian, M. 2011, MNRAS, 418, 759
 
 \bibitem[{{Renaud} \& {Gieles}(2015)}]{renaud15} Renaud, F. \& Gieles, M. 2015, MNRAS, 448, 341
  


 \bibitem[{{Romanowsky} \& {Kochanek}(2001)}]{romanowsky01} Romanowsky, A.J. \& Kochanek, C.S. 2001, ApJ, 553, 722

 \bibitem[{{Romanowsky} {et~al.}(2012)}]{romanowsky12} Romanowsky, A.J., Strader, J.,  Brodie, J.P., Mihos, J.C., Spitler, L.R, Forbes, D.A., Foster, C., Arnold, J.A. 2012, ApJ, 748, 29
  
 
   \bibitem[{{Sippel} {et~al.}(2012)}]{sippel12} Sippel, A.C., Hurley, J.R., Madrid, J.P., Harris, W.E. 2012, MNRAS, 427, 167
 
\bibitem[{{Schiminovich} {et~al.}(1994)}]{schiminovich94}  Schiminovich, D., van Gorkum, J. H., van der Hulst, J. M., \& Kasow, S. 1994, ApJ, 423, L101
  
  \bibitem[{{Schuberth} {et~al.}(2010)}]{schuberth10} Schuberth, Y., Richtler, T., Hilker, M., Dirsch, B., Bassin, L.P., Romanowsky, A.J., Infante, L. 2010, A\&A, 512, 52
 
  
 \bibitem[{{Schulman} {et~al.}(2012)}]{schulman12}Schulman, R. D., Glebbeek, E., Sills, A. 2012, MNRAS, 420, 651
 


\bibitem[{{Spitler} {et~al.}(2006)}]{spitler06} Spitler, L.R., Larsen, S.S., Strader, J., Brodie, J.P., Forbes, D.A., Beasley, M.A. 2006, AJ, 132, 1593


\bibitem[{{Strader} {et~al.}(2011)}]{strader11} Strader, J., Romanowsky, A.J., Brodie, J.P., Spitler, L.R., Beasley, M.A., Arnold, J.A., Tamura, N., Sharples, R.M., Arimoto, N. 2011,APJS, 197, 33 

\bibitem[{{Strader} {et~al.}(2012)}]{strader12} Strader, J., Fabbiano, G., Luo, B., Kim, D., Brodie, J.P., Fragos, T., Gallagher, J.S., Kalogera, V., King, A., Zezas, A. 2012, ApJ, 760, 87

 \bibitem[{{Tal} {et~al.}(2009)}]{tal09} Tal, T., van Dokkum, P. G., Nelan, J., Bezanson, R. 2009, AJ, 138, 1417


  \bibitem[{Tonini (2013)}]{tonini13} Tonini, C. 2013, ApJ, 762, 39
  
 \bibitem[{{Tonry} {et~al.}(2001)}]{tonry01} Tonry, J. L., Dressler, A., Blakeslee, J. P., Ajhar, E.A., Fletcher, A.B., Luppino, G.A., Metzger, M.R., Moore, C.B. 2001, ApJ, 546, 681

\bibitem[{{Tremaine} {et~al.}(1994)}]{tremaine94} Tremaine, S., Richstone, D.O., Byun, Y.-L., Dressler, A., Faber, S. M., Grillmair, C., Kormendy, J. and Lauer, T. R. 1994, AJ, 107, 634


\bibitem[{{Usher} {et~al.}(2013)}]{usher13} Usher, C., Forbes, D.A., Spitler, L.R., Brodie, J.P., Romanowsky, A.J., Strader, J., Woodley, K.A. 2013, MNRAS, 436, 1172


 \bibitem[{{van den Bergh} (2003)}]{vdbergh03} van den Bergh, S. 2003, ApJ, 590, 797


 \bibitem[{{van Dokkum} (2003)}]{vandokkum03} van Dokkum, P. G. 2005, AJ, 130, 2647

 \bibitem[{{Vesperini} {et~al.}(2003)}]{vesperini03} Vesperini, E., Zepf, S. E., Kundu, A., Ashman, K. M. 2003, ApJ, 593, 760 
 
  \bibitem[{von Hoerner (1957)}]{vonhoerner57} von Hoerner, S. 1957, ApJ, 125, 451
 
 \bibitem[{{Webb} {et~al.}(2012)}]{webb12} Webb, J.J, Sills, A., Harris, W.E. 2012, ApJ, 746, 93

  \bibitem[{{Webb} {et~al.}(2013a)}]{webb13a} Webb, J.J., Harris, W.E., Sills, A., Hurley, J.R. 2013a, ApJ, 764, 124
 
  \bibitem[{{Webb} {et~al.}(2013b)}]{webb13b} Webb, J.J, Sills, A., Harris, W.E. 2013b, ApJ, 779, 94
  
    \bibitem[{{Webb} {et~al.}(2014)}]{webb14} Webb, J.J., Sills, A., Harris, W.E., Hurley, J.R. 2014, MNRAS, 445, 1048
  
\bibitem[{{Weijmans} {et~al.}(2009)}]{weijmans09} Weijmans, A., Cappellari, M., Bacon, R., de Zeeuw, P. T., Emsellem, E., Falcon-Barroso, J., Kuntschner, H., McDermid, R. M., van den Bosch, R. C. E., and van de Ven, G., 2009, MNRAS, 398, 561

 
  \bibitem[{{Woodley} {et~al.}(2007)}]{woodley07} Woodley, K.A., Harris, W.E., Beasley, M.A., Peng, E.W., Bridges, T.J., Forbes, D.A., Harris, G.L.H. 2007, AJ, 134, 494
 
 \bibitem[{{Woodley} \& {G\'{o}mez}(2010a)}]{woodley10a} Woodley, K.A. \& G\'{o}mez, M. 2010a, PASA, 27, 379
 \bibitem[{{Woodley} {et~al.}(2010b)}]{woodley10b} Woodley, K.A., G\'{o}mez, M., Harris, W.E., Geisler, D., Harris, G.L.H. 2010b, AJ, 139, 1871
 \bibitem[{{Woodley} {et~al.}(2010c)}]{woodley10c} Woodley, K.A., Harris, W.E., Puzia, T.H., G\'{o}mez, M., Harris, G.L.H., Geisler, D. 2010c, ApJ, 708, 1335

 \bibitem[{{Woodley}(2012)}]{woodley12} Woodley, K. 2012, AAS Meeting $\# 220$, $\# 438.07$

 \bibitem[{{Xu} {et~al.}(2014)}]{xu14} Wu, X., Gerhard, O., Naab, T., Oser, L., Martinez-Valpuesta, I., Hilz, M., Churazov, E., Lyskova, N. 2014, MNRAS, 438, 2701
 
\bibitem[{{Zait}, {Hoffman}, \& {Shlosman}(2008)}]{zait08} Zait, A., Hoffman, Y. \& Shlosman, I. 2008, ApJ, 682, 835

\bibitem[{{Zepf} \& {Ashman}(1993)}]{zepf93} Zepf, S. E. \& Ashman, K. M. 1993, MNRAS, 264, 611

 \bibitem[{{Zhu} {et~al.}(2014)}]{zhu14} Zhu, L., Long, R.J., Mao,S., Peng, E.W., Liu, C., Caldwell, N., Li, B., Blakeslee, J.P, C\^{o}t\'{e}, P., Cuillandre, J.C., Durrell, P., Emsellem E., Ferrarese, L., Gwyn, S., J\'{o}rdan,  A., Lan\c{c}on, A., Mei, S., Mu$\tilde{n}$oz, R., Puzia, T. 2014, ApJ, 729, 59 


\end{thebibliography}
\end{document}